\documentclass[review,a4paper]{elsarticle}
\makeatletter
\def\ps@pprintTitle{%
 \let\@oddhead\@empty
 \let\@evenhead\@empty
 \def\@oddfoot{\centerline{\thepage}}%
 \let\@evenfoot\@oddfoot}

\usepackage{lineno,hyperref}
\modulolinenumbers[5]

\usepackage{bm}
\usepackage{mathtools}
\usepackage{latexsym}
\usepackage{soul}
\usepackage{siunitx}
\usepackage{afterpage}
\usepackage{geometry}
\usepackage{pdfpages}
\usepackage{textgreek}
\usepackage{graphicx,dblfloatfix} 
\usepackage{natbib} 
\usepackage{lineno}
\usepackage[utf8]{inputenc} 
\usepackage{amsmath}
\usepackage{amsthm}
\usepackage{amssymb}
\usepackage{color}
\usepackage{xcolor}
\usepackage{subcaption}
\usepackage{comment}
\usepackage{float}
\usepackage{a4wide}
\usepackage{booktabs}
\usepackage{multirow}
\usepackage{tabularx}
\usepackage{soul}

\newcolumntype{L}{>{\raggedright\arraybackslash}X}

\newcommand{\vgrad}{\boldsymbol{\nabla}}

\newcommand{\vdiver}{\operatorname{\mathbf{div}}\,}
\newcommand{\diver}{{\operatorname{div}}\,}

\journal{Carbon}

 \usepackage{numcompress}

\begin{document}

\begin{frontmatter}

\title{Quantifying stress distribution in ultra-large graphene drums through mode shape imaging}

\author[1staddress]{Ali Sarafraz}
\author[1staddress]{Hanqing Liu}
\author[4thaddress]{Katarina Cvetanović}
\author[4thaddress]{Marko Spasenović}
\author[3rdaddress]{Sten Vollebregt}
\author[3rdaddress]{Tomas Manzaneque Garcia}
\author[1staddress,2ndaddress]{Peter G. Steeneken}
\author[1staddress]{Farbod Alijani}

\author[1staddress]{Gerard J. Verbiest\corref{mycorrespondingauthor2}}
\cortext[mycorrespondingauthor2]{Corresponding author. Email:G.Verbiest@tudelft.nl}

\address[1staddress]{Department of Precision and Microsystems Engineering, Faculty of Mechanical, Maritime and Materials
Engineering, Delft University of Technology, 2628 CD, Delft, The Netherlands.}
\address[4thaddress]{Institute of Chemistry, Technology and Metallurgy, University of Belgrade, Njegoševa 12, 11000 Belgrade, Serbia}
\address[3rdaddress]{Department of Microelectronics, Delft University of Technology, Mekelweg 4, 2628 CD, Delft, The Netherlands}
\address[2ndaddress]{Kavli Institute of Nanoscience, Faculty of Applied Sciences, Delft University of Technology, 2628 CJ, Delft, The
Netherlands}

\begin{abstract}
Suspended drums made of 2D materials hold potential for sensing applications. However, the industrialization of these applications is hindered by significant device-to-device variations presumably caused by non-uniform stress distributions induced by the fabrication process. Here we introduce a new methodology to determine the stress distribution from their mechanical resonance frequencies and corresponding mode shapes as measured by a laser Doppler vibrometer (LDV).
To avoid limitations posed by the optical resolution of the LDV, we leverage a unique manufacturing process to create ultra-large graphene drums with diameters of up to $1000~\mu\si{m}$. 
We solve the inverse problem of a Föppl--von Kármán plate model by an iterative procedure to obtain the stress distribution within the drums from the experimental data. Our results show that the generally used uniform pre-tension assumption overestimates the pre-stress value, exceeding the averaged stress obtained by more than 47\%. Moreover, it is is found that the reconstructed stress distributions are bi-axial, which likely originates from the transfer process.
The introduced metholodogy allows one to estimate the tension distribution in drum resonators from their mechanical response and thereby paves the way for linking the used fabrication processes to the resulting device performance.

\textbf{keywords:} Stress distribution, graphene resonator, resonance, mode shapes, frequency split, stress quantification
\end{abstract}

\end{frontmatter}

\section{Introduction}
The exceptional mechanical properties of suspended two-dimensional (2D) materials such as graphene make them ideal materials for applications such as force, mass, and sound sensing \cite{lemme2020nanoelectromechanical,yildirim2020towards,hu2020resonant, Todorović2015}. Their unique opto- and electromechanical coupling has enabled studies into phase transitions \cite{vsivskins2020magnetic}, heat transport \cite{dolleman2017optomechanics, liu2022tension} and even measuring the biological forces of micro-organisms~\cite{Roslon2022}. However, the industrial realization of 2D materials is currently hindered by significant device-to-device variations observed in practice~\cite{Ferrari2023}. Addressing this variability is crucial for enhancing the reproducibility and reliability of 2D material based devices. A key factor contributing to this variability is the built-in stress arising from the fabrication process \cite{Ferrari2023, Akinwande2017, Vozmediano2010}.


During the transfer of 2D materials onto target substrates, non-uniform stress distributions inevitably occur~\cite{chen2019wrinkling, deng2016wrinkled}, resulting in surface defects like wrinkles in the fabricated drums~\cite{steeneken2021dynamics, Sarafraz2021, Gornyi2016}. Currently, Raman spectroscopy and Atomic Force Microscopy (AFM) are the methods of choice for analyzing the stress distribution in 2D materials \cite{paillet2018graphene,dai2019strain, colangelo2019mapping}. Raman spectroscopy monitors the strain-sensitive position of Raman active phonon modes~\cite{ferralis2010probing}. However, due to its limited spatial and strain resolution, it only provides a relatively coarse strain measurement, making it less suitable for the quantification of the initial stress in suspended 2D materials. AFM, in contrast, is a contact-based technique that is challenging to perform and applies force to the membrane during measurement. This force has the potential to affect the morphology and distribution of tension in the membrane, despite its effectiveness in quantifying stress distribution. Consequently, development of new non-contact methods that can determine the stress distribution in 2D membranes is highly desirable.

In this paper, we propose a novel methodology to quantify the stress distribution of ultra-thin suspended drums by using nanomechanical resonances and their mode shapes. We use Laser Doppler Vibrometry (LDV)~\cite{Castellini2006} to measure graphene drums, with large diameters from $60~ \mu\si{m}$ to $1000~ \mu\si{ m}$, capturing their dynamics with picometre-amplitude resolution. 
Subsequently, we create an analytical model to calculate resonance frequencies and mode shapes. We then follow a reverse-path by using experimental data to predict both in-plane and out-of-plane displacements, as well as the stress distribution of the 2D drum. Our results highlight that 2D material drums are not uniformly tensioned which challenges the current methodologies for estimating the built-in stress of these drums~\cite{colangelo2019mapping, ferralis2010probing}. The presented methodology allows fabrication techniques to be optimized for improving the uniformity and reproducibility of stress distributions, thus improving yield and performance of sensing applications based on suspended 2D material membranes.

\section{Experiments}
\subsection{Fabrication method}

\begin{figure}
	\includegraphics[width=1\linewidth]{"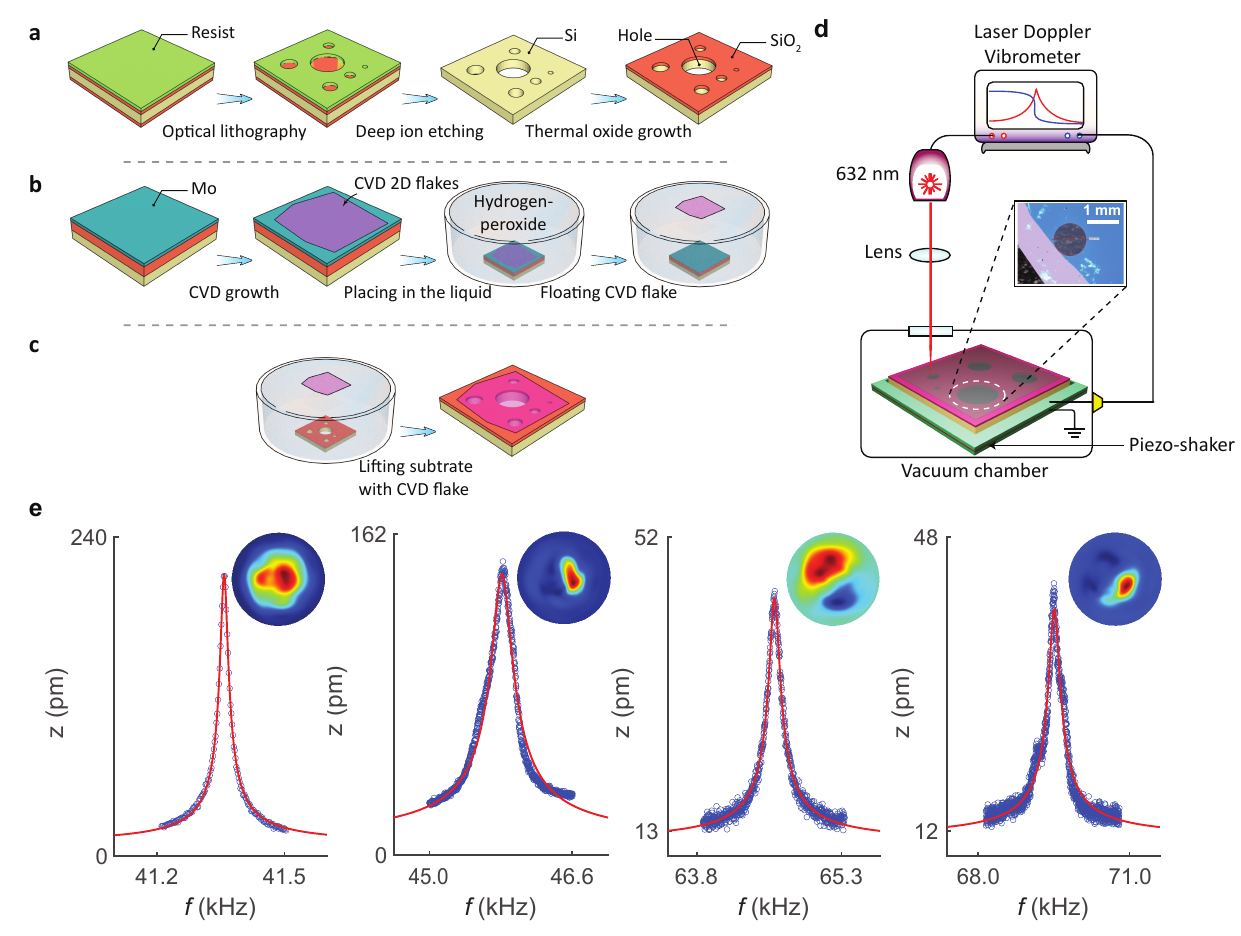"}
    \centering
	\caption{Fabrication and vibration measurement of graphene drums. (a) Fabrication process of SiO$_2$/Si substrate with etched holes. (b) Growth and exfoliation of large-scale CVD graphene flake. (c) Wet transfer method to suspend CVD graphene on substrate. (d) Schematic of measurement setup comprising a MSA400 Polytec Laser Doppler Vibrometer (LDV) for detection and read-out. The sample is placed in the vacuum chamber and driven by a piezo shaker. The inset: optical image of device D1; purple region, Si/SiO$_2$ substrate; blue region, supported graphene; transparent region, suspended graphene. (e) The first four resonance of device D1; the damped linear harmonic oscillator fit is shown by red line.}
    \label{fig:experiment}
\end{figure}

The fabrication procedure employed for manufacturing the devices is illustrated in figures~\ref{fig:experiment}(a)-(c). As figure~\ref{fig:experiment}(a) shows, we first prepare a Si (100) target substrate containing through holes  etched using deep reactive ion etching with diameters ranging from $60~ \mu\si{m}$ to $1000~ \mu\si{ m}$.
Next, we deposit multi-layer graphene using chemical vapor deposition (CVD) on a thin-film Mo catalyst, as shown in figure~\ref{fig:experiment}(b). 
The final stage of the fabrication procedure is the transfer of CVD-grown graphene from the growth substrate to the target substrate, executed through a wet transfer process, as depicted in figure~\ref{fig:experiment}(c). 

The fabrication process resulted in the production of a set of 16 unique devices, denoted as D1 to D16, which were spread among different chips. For a comprehensive description of the fabrication method, please refer to supplementary information S1. The drums had a range of radii (\(R\)) from 61 to 1032~$\mu$m. We measured the thickness (\(h\)) of the CVD graphene on all chips using an atomic force microscopy (AFM) and found that $h$ ranges from 7 to 13.8 nm (see supplementary information S2). 

\subsection{Measurement methodology and results}

To probe nanomechanical vibrations of the devices, we use a piezo shaker to drive the drums into resonance and a Polytec MSA400 Laser Doppler Vibrometry (LDV) system to read-out their velocity in the out-of-plane direction (see figure~\ref{fig:experiment}(d)). The LDV actuates the piezo shaker at a specific frequency $f$ while simultaneously recording the position-dependent displacement $z_f$ of the device using a 632 nm He-Ne laser. All measurements are conducted at room temperature inside a vacuum chamber at 10$^{-4}$~mbar.

\begin{figure}
	\includegraphics[width=1\linewidth]{"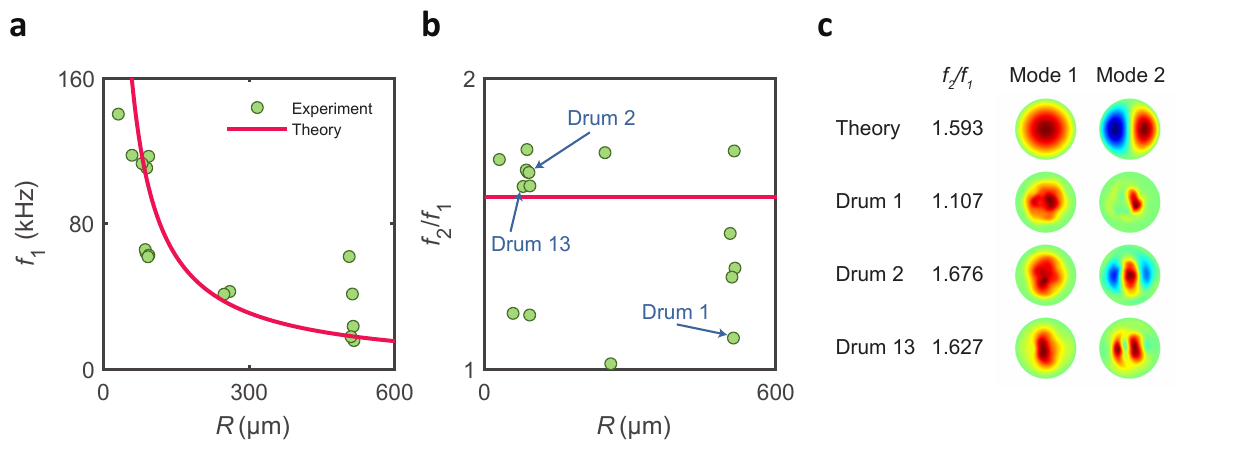"}
	\caption{Comparison of experimental results with uniform pre-tension. (a) The relationship between the fundamental frequency $f_1$ and the drum radius $R$ for devices D1$-$D16 is observed to exhibit an inverse trend. Resonance frequency estimation using equation~(\ref{eq:simpleUnifrom}) is shown in red, by assuming thickness $h=15$~nm and pre-tension $n_0=0.02$~N/m. (b) The ratio of the second to the first frequency ($f_2/f_1$) as a function of $R$ for all studied devices. The theoretical value of $f_2/f_1$  for drums having uniform pre-tension is shown by red line. (c) The first and second mode shapes and the corresponding frequency ratio of three of the measured devices and their comparison to theoretical estimates based on equation~(\ref{eq:simpleUnifrom}). All mode shapes are normalized with respect to the maximum value of their displacement, and the colorbar ranges between -1 and 1.}
    \label{fig:mode_split}
\end{figure}

Figure~\ref{fig:experiment}(e) displays the measured first four resonances of device D1. To extract the resonances, we employ a fitting procedure based on a linear harmonic oscillator model (represented by the red line in figure~\ref{fig:experiment}(e)). The frequency characteristics of the first mode ($f_1=\frac{\omega_{1}}{2\pi}$) in devices D1-D16, as a function of the radius $R$, are depicted in figure~\ref{fig:mode_split}(a), aligning with the behavior observed in circular drums' fundamental frequencies, as previously reported~\cite{castellanos2013single}. The relationship governing the resonance frequencies is described by the equation
\begin{equation}
\label{eq:simpleUnifrom}
f_{i} = \frac{\gamma_{i}}{2 \pi R}\sqrt{\frac{n_0}{\rho h}},
\end{equation}    
where $\gamma_{i}$ represents a constant, $n_0$ denotes the pre-tension applied to the drum, $\rho$ signifies the mass density, and $h$ stands for the thickness of the drum. Theory gives $\gamma_{1}=2.4048$, with higher resonance frequencies corresponding to increased values of $\gamma_i$ relative to $\gamma_1$. Since $R$ and $h$ are known for our samples from optical microscopy and AFM measurements, we can extract $n_0$ of the devices using the first resonance frequency $f_1$ by utilizing equation~(\ref{eq:simpleUnifrom}). We note that the obtained $n_0$, varies from $2\times 10^{-3}$ to $3\times 10^{-1}$~N/m, which is comparable to the values reported in the literature \cite{chen2009performance,zande2010large}. The corresponding strain $\varepsilon_0$ extracted from $n_0=\sigma_0 h=Eh\varepsilon_0/(1-\nu)$ is below 0.0013$\%$ for all our devices, which is much lower than the resolution limits of Raman spectroscopy  \cite{mohiuddin2009uniaxial}.




In figure~\ref{fig:mode_split}(b), we plot the ratio between the second and the first resonance frequency of 16 fabricated devices. Included in the figure we also show the $f_2/f_1$ obtained analytically for circular drums (red line). We note that the experimental values of $f_2/f_1$ significantly deviate from this red line with a minimum ratio of 1.019 and a maximum ratio of 1.754. To gain insight into these deviations, in figure ~\ref{fig:mode_split}(c) we show the experimental mode shapes, as determined by the MSA-400 vibrometer (figure~\ref{fig:experiment}(d)), for three of the devices. We note clear differences between the second theoretical mode shape of a circular drum and the experimental mode shape, which emphasizes the substantial influence of non-uniform stress on the dynamics of these drums. In addition, we analyse the ratios $f_3/f_1$ and $f_4/f_1$ for a specific subset of our drums (see supplementary information S3), in which we also observe a significant difference between the experimental findings and the theoretical predictions based on the assumption of uniform pre-tension.

In addition, it has been theoretically predicted that when a stress distribution is uniform, it results in the emergence of asymmetric mode shapes that are defined by $n$ nodal lines rotated by $2\pi/n$ relative to each other, and possess equal resonance frequencies~\cite{reddy2006}. Nevertheless, as the level of stress non-uniformity increases, these mode shapes undergo substantial changes resulting in a loss of resemblance between them. Consequently, the non-uniformity in tension distribution has a substantial effect on the mode-shapes, and therefore these mode-shapes contain important information on the stress distribution. In the subsequent section, we introduce a method to deduce the non-uniform tension distribution based on the experimentally acquired mode shapes and resonance frequencies.



\section{Quantifying tension distribution}
\subsection{Theory}

To analyze the effect of stress distributions on the mode-shapes of the drums, we employ a circular plate model characterized by radius $R$ and thickness $h$. This model assumes the material to be homogeneous and isotropic, having a density $\rho$, Young's modulus $E$, and Poisson's ratio $\nu$. The governing equations are expressed in cylindrical coordinates ($r$, $\theta$, $z$), with $r$ representing the radial, $\theta$ the azimuthal, and $z$ the transverse coordinate. The equations governing transverse and in-plane motions, derived through Hamilton's principle, are given by~\cite{amabili2008nonlinear}:
\begin{subequations}
\label{eq:sysNLPlak}
\begin{align}
\rho h \ddot{w} + D{\nabla ^4}w - \diver(\bm{N}\vgrad w) &= 0,  \label{eq:sysNLPlakA}\\
\rho h \ddot{\bm{u}} - \vdiver\bm{N} & = 0\label{eq:sysNLPlakB}, 
\end{align}
\end{subequations} 
where
\begin{equation}
\label{eq:Neps}
\begin{aligned}
\bm{N}=[Eh/(1- \nu^2)]\left[(1-\nu)\bm{\epsilon}+\nu\text{tr}(\bm{\epsilon})\bm{I}\right], \\
\bm{\epsilon}=\frac{1}{2}\left(\nabla\bm{u}+\nabla^T\bm{u} + \nabla w\otimes\nabla w\right).
\end{aligned}
\end{equation}
In the equations above, $\bm{u}=[u; v]$, with $u$ and $v$ denoting the radial and azimuthal displacement fields, while $w$ represents the transverse displacement field. Additionally, $\nabla^4 w$, $\nabla w$, and $\nabla \bm{u}$ denote the biharmonic operator applied to the scalar field $w$, the vector gradient of the scalar field $w$, and the tensor gradient of the vector field $\bm{u}$, respectively. $\vdiver \bm{N}$ is vector divergence of the tensor field $\bm{N}$. Furthermore, $\nabla w\otimes\nabla w$ corresponds to the tensor product between vectors $\nabla w$ and $\nabla w$. An overdot indicates differentiation with respect to time, and $D = \frac{E{h^3}}{{12(1 - {\nu ^2})}}$ denotes the bending rigidity. It is noteworthy that, as per the notation presented here, the strain tensor $\bm{\epsilon}$ and stress resultant tensor $\bm{N}$ can be identified as second-order tensors in a two-dimensional framework.

In practice, fabricated drums may exhibit deformations that deviate from the conventional assumption of uniform radial deformation, often associated with uniform pre-tension. Consequently, when these drums undergo transverse dynamic actuation, their displacement fields comprise both static and dynamic components. The static deformation originates from the pre-actuation displacement history, while the dynamic component represents the displacement induced by the actuation process. To gain a comprehensive understanding of the mechanical response in such situations, it becomes essential to incorporate both static and dynamic displacements within the overall displacement field~\cite{Sarafraz2023, Sajadi2017,Li2010}. However, the substantial difference in magnitude between in-plane and transverse inertia necessitates the exclusion of dynamic deformation in the in-plane direction~\cite{reddy2006}. Therefore, we assume $\bm{u}=\bm{u}_s$, but $w = w_s + w_d$, where the subscript $s$ refers to static components and the subscript $d$ represents dynamic deformations. It is important to acknowledge that the value of $w_s$ is typically non-zero, since the membranes may exhibit wrinkling or bulging following their fabrication process.

To capture the vibrational response (\(w_d\)) of these drums, we conduct a modal analysis using equation~(\ref{eq:sysNLPlakA}) centered around the statically deformed configuration \((u_s,v_s,w_s)\). However, since equation~(\ref{eq:sysNLPlakA}) involves \(\bm{N}\) and is not expressed in terms of displacement fields, we initially reformulate the equation in the context of static and dynamic displacement fields. The detailed derivation procedure for this can be found in supplementary information S4. Next , we assume the dynamic transverse deformation $w_d$ to be harmonic and express it as $w_d(r,\theta,t) = w_0^d\varphi(r,\theta)\exp(i\omega t)$,
where \(w_d^0\) represents the maximum spatial amplitude of the drum at time \(t=0\), \(\varphi(r,\theta)\) denotes the mode shape normalized with respect to maximum displacement, and \(\omega\) is the corresponding resonance frequency. Next, we make the equations dimensionless (see supplementary information S4) and discretize them over a mesh with \(N = 161\) nodes in the radial direction and \(M = 91\) nodes in the azimuthal direction (see supplementary information S5), which leads to the compact form of the transverse governing equation
\begin{equation}
\label{eq:discrete_matrix_form}
{{\bf{D}}_{\bf{U}}}{U_{i,j}} + {{\bf{D}}_{\bf{V}}}{V_{i,j}} + \sum\limits_k {\left( {{\bf{\bar D}}_{\bf{W}}^kW_{i,j}} \right) \cdot \left( {{\bf{\bar{\bar{D}}}}_{\bf{W}}^kW_{i,j}} \right)}  = \left( {{\bar{\omega} ^2}I - {{\bf{D}}_{\bf{W}}}} \right){\Phi_{i,j}},
\end{equation}
where \(U_{i,j}\), \(V_{i,j}\), and \(W_{i,j}\) represent the unknown static deformations at spatial node \((i,j)\). Additionally, \(\Phi_{i,j}\) is the given (or known) normalized mode shape extracted from the experiments, \(\bar{\omega}\) denotes the corresponding non-dimensional resonance frequency, and $I$ is the identity matrix. Moreover, the matrices \({{\bf{D}}_{\bf{U}}}\), \({{\bf{D}}_{\bf{V}}}\), \({{\bf{D}}_{\bf{W}}}\), \({{\bf{\bar D}}_{\bf{W}}^k}\), and \({{\bf{\bar{\bar{D}}}}_{\bf{W}}^k}\) denote linear differential operators dependent on the mode shapes and discretization weighting coefficients. Comprehensive details regarding this step can be found in supplementary information S5.

Unlike the conventional modal analysis, where predefined static deformations \(U_{i,j}\), \(V_{i,j}\), and \(W_{i,j}\) are used to deduce resonance frequencies \(\bar{\omega}\) and mode shapes \(\Phi_{i,j}\), in our approach, we follow a reverse-path, and deduce these deformations from measured resonance frequencies and mode shapes. Given the existence of three unknown displacement fields, specifically \(U_{i,j}\), \(V_{i,j}\), and \(W_{i,j}\) (\(3 \times M \times N\) unknowns), the extraction of these displacement fields necessitates a minimum of \(3 \times M \times N\) equations. This underscores the significance of having no fewer than three sets of mode shapes \((\Phi_{i,j}^1, \Phi_{i,j}^2, \Phi_{i,j}^3)\), along with their corresponding non-dimensional frequencies \((\bar{\omega}_1, \bar{\omega}_2, \bar{\omega}_3)\) for estimating the built-in stress, which collectively provide \(3 \times M \times N\) equations across the mesh (see equation~(\ref{eq:discrete_matrix_form})). However, in practice, an additional mode shape becomes a crucial requirement. The underlying rationale for this is rooted in the observation that equations linked to boundary nodes yield a trivial \(0=0\) relationship. Consequently, introducing supplementary equations is necessary to fulfill the requisite rank of the algebraic equation system. In pursuit of accurately determining stress distributions from experimental mode shapes and frequencies, it thus becomes essential to consider at least four mode shapes alongside their corresponding resonance frequencies.

To validate our numerical methodology and equations, we performed finite element simulations on a flat circular plate characterized by a predefined non-uniform stress distribution. The computed mode shapes and resonance frequencies were then employed to reconstruct the stress distribution using the methodology we have introduced (for detailed discussion, see supplementary information S6).

\begin{figure}[h!]
	\includegraphics[width=1\linewidth]{"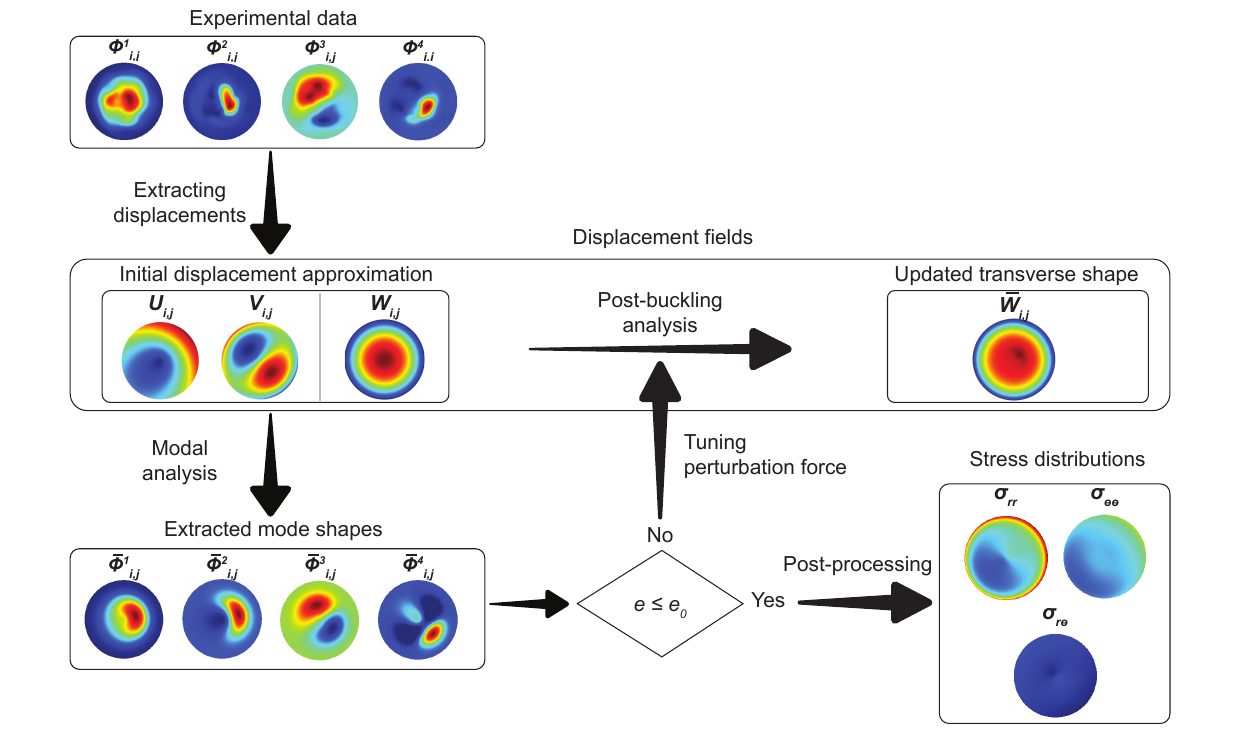"}
	\caption{The flowchart for deriving stress distributions from experimental mode shapes and resonance frequencies. Assuming a parabolic deformation in the transverse direction, the experimental mode shapes are first employed to extract initial displacement fields. Then a modal analysis is performed. If the extracted mode shapes do not match the experimental ones sufficiently ($e>e_0$), the perturbation force is tuned and using a post-buckling analysis, the transverse shape is modified. After that, using in-plane displacement fields and the updated transverse displacement, we again obtain theoretical mode shapes and resonance frequencies. If the mode shapes do not satisfy the criterion~(\ref{eq:error_criteria}), we recalculate the transverse shape by adjusting the perturbation force to the post-buckling step. From there we return to the modal analysis. When the convergence criterion is met, the obtained displacement fields are used to calculate stress distributions. Here, $i = 1,2, \ldots ,N$, and $j = 1,2, \ldots ,M$ denotes the node number in radial, and azimuthal directions.}
    \label{fig:flowchart}
\end{figure}

By simultaneously using the governing equation~(\ref{eq:discrete_matrix_form}) for a minimum of four mode shapes and the respective resonance frequencies, it becomes possible to determine the static displacement fields and the associated stress distribution fields. The flowchart presented in figure~\ref{fig:flowchart} explains the sequential approach for obtaining stress distributions from experimental mode shapes and frequencies. The technique commences by fitting a surface to the experimental mode shapes. This is important as equation~(\ref{eq:discrete_matrix_form}) involves derivative operators and any non-smoothness and noise in experimental mode shapes leads to numerical inaccuracies. In order to guarantee the compliance of boundary conditions, we utilize the mode shapes of a uniformly-tensioned plate as the basis for our fitting function (see supplementary information S7).

Subsequently, utilizing the smoothed mode shapes, we aim to extract the static displacement fields. However, due to the nonlinearity of equation (\ref{eq:discrete_matrix_form}) with respect to $W_{i,j}$, extracting the solution without a suitable initial approximation poses a challenge. As a possible solution, we assume a parabolic form for the static transverse displacement, characterized by an undetermined deflection amplitude at the center of the drum (\(W_{i,j} = W_0(1-\bar{R}_{i,j}^2)\)), where \(W_0\) signifies the deflection at the center, and $\bar{R}_{i,j}$ represents the $\bar{r}$-coordinate of node $(i,j)$. By solving equation~(\ref{eq:discrete_matrix_form}) using the experimentally acquired mode shapes \(\Phi_{i,j}^1\), \(\Phi_{i,j}^2\), \(\Phi_{i,j}^3\), and \(\Phi_{i,j}^4\), we can determine the unknowns \(U_{i,j}\), \(V_{i,j}\), and \(W_0\). It is crucial to recognize that due to the influence of noise and measurement inaccuracies, achieving 100\% accuracy in solving for displacements is unattainable. Therefore, employing a least-squares method becomes necessary. This method enhances accuracy by incorporating more equations, namely additional mode shapes and frequencies, into the solution process.

It is imperative to acknowledge that assuming an axisymmetric parabolic deflection for the transverse displacement has inherent limitations. The presence of non-uniform displacements in the studied membranes might lead to static deformation and the creation of a buckling pattern, owing to their ultra-thin nature. Due to experimental and numerical inaccuracies, the displacement fields that are obtained from solving equation (\ref{eq:discrete_matrix_form}) don't always meet the requirements of the in-plane equation (\ref{eq:sysNLPlakB}). Therefore, it is crucial to identify a stable out-of-plane configuration that fulfills equation~(\ref{eq:sysNLPlak}). In response to potential static transverse asymmetries and to rectify our initial assumption of parabolic static deformation, we perform a post-buckling analysis. This analysis utilizes the in-plane displacement fields \(U_{i,j}\), \(V_{i,j}\), and \(W_{i,j}\), which are obtained from the experimental mode shapes to update transverse displacement $W_{i,j}$ to $\bar{W}_{i,j}$. In the context of this post-buckling analysis, a minor perturbing uniform transverse force is introduced, which serves to update the drum's transverse shape and accounts for its nonlinear behavior~\cite{Changguo2007}.

In order to perform the post-buckling analysis, it is recommended to utilize equations~(\ref{eq:sysNLPlak}) or alternatively, incorporate the in-plane displacement fields ($U_{i,j}$ and $V_{i,j}$) into a finite element method (FEM) software that is capable of handling nonlinear structural analysis. This will enable an analysis of the post-buckling response of the drum and the establishment of its modified transverse shape $\bar{W}_{i,j}$. It is crucial to highlight that as a result of the non-uniform characteristics of $U_{i,j}$ and $V_{i,j}$, the post-buckling analysis results in an asymmetric transverse shape $\bar{W}_{i,j}$ that deviates from the axisymmetric shape $W_{i,j}$.

To ensure the accurate computation of both in-plane and transverse displacement fields, we conduct a modal analysis to extract mode shapes and resonance frequencies from the static displacements (\(U_{i,j}\), \(V_{i,j}\), and \(\bar{W}_{i,j}\)). In order to quantitatively assess the fidelity of the reconstructed mode shapes compared to their experimental counterparts, we employ an error metric denoted by \(e\). This metric \(e\) characterizes the average spatial deviation between the reconstructed and experimental mode shapes and is defined as follows
\begin{equation}
\label{eq:error_criteria}
e = \frac{1}{{\bar N}}\sum\limits_{n = 1}^{\bar N} {{{\left[ {\frac{{\int_0^{2\pi } {\int_0^1 {{{\left( {{{\bar \varphi }_n} - {\varphi _n}} \right)}^2}r{\rm{d}}r{\rm{d}}\theta } } }}{{\int_0^{2\pi } {\int_0^1 {{{\left( {{\varphi _n}} \right)}^2}r{\rm{d}}r{\rm{d}}\theta } } }}} \right]}^{{1 \mathord{\left/
 {\vphantom {1 2}} \right.
 \kern-\nulldelimiterspace} 2}}}}  \le {e_0},
\end{equation}
where \(e_0\) serves as an error threshold. In this equation, \({{\bar \varphi }_n}\) is the \(n\)-th normalized mode shape obtained from modal analysis, \({{\varphi _n}}\) signifies the \(n\)-th normalized experimental mode shape, and \(\bar{N}\) represents the total number of mode shapes used in the fitting procedure. It's important to note that both \({{\bar \varphi }_n}\) and \({{\varphi _n}}\) are the continuous forms of \(\Phi _{i,j}^n\) and \(\bar{\Phi} _{i,j}^n\), respectively.

If the criterion (\ref{eq:error_criteria}) is satisfied (\(e \le e_0\)), the solution is considered to be converged. Conversely, if \(e > e_0\), the post-buckled configuration is re-calculated with a new perturbation in the transverse direction, leading to the acquisition of an updated transverse mode shape. This iterative process continues until the convergence criterion is met.

To determine an appropriate value for \(e_0\), we initiate the iterative process without applying any perturbation force and gradually increase it step-by-step. The observed trend reveals an initial decrease in the error metric \(e\) as the perturbation force rises until it reaches the minimum value \(e_0\) at a perturbation force of \(\delta p_m\). Beyond this point, further increments in the perturbation force result in an increase in the error. Consequently, the minimum achievable error for each set of experiments corresponds to \(e_0\), which varies among different drums. For instance, device D1 exhibits an error threshold of \(e_0 \simeq {\rm{0}}{\rm{.19}}\). A more detailed and comprehensive discussion regarding the determination of \(e_0\) can be found in supplementary information S8.

Once the solution has converged, the numerical displacement field effectively approximates the experimental displacement field, which enables us to compute the strain field and subsequently derive the stresses within the drum's mid-plane using equations (S2) and (S3). For a more comprehensive overview of the described procedure, including a detailed flowchart, please consult the supplementary information S7.

\subsection{Fitting results}

As elaborated in the previous section, the numerical procedure necessitates an initial assumption of a parabolic transverse displacement field. However, this assumption does not universally apply to all manufactured devices. Some of the manufactured devices exhibit significant complex wrinkling patterns that deviate noticeably from the parabolic approximation. Consequently, the proposed method is not applicable to drums that deviate from this assumption.

\begin{figure}
	\includegraphics[width=1\linewidth]{"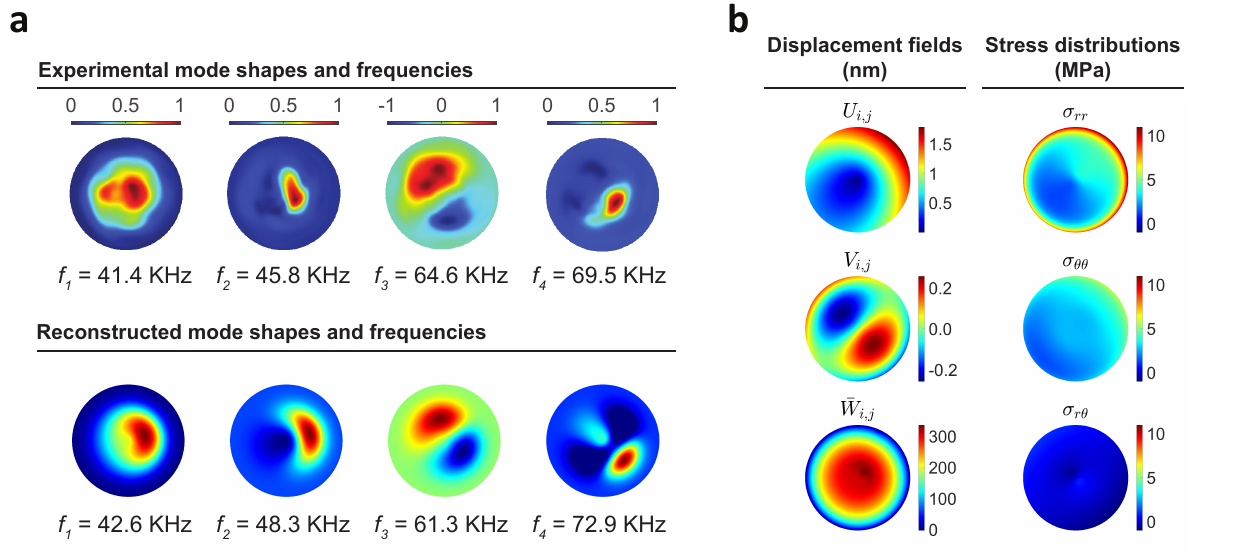"}
	\caption{(a) Reconstruction of stress distributions based on experimental mode shapes  for device D1. As described in the main text, the experimental mode shape and frequencies are utilised to derive displacement fields and stress distributions. For the purpose of determining the validity of the results, the theoretical mode shapes are reconstructed using the displacement field. (b) Stress distribution predicted by current method in comparison with predictions of Raman spectroscopy. As can be seen, the Raman's results show a uniform stress distribution as opposed by our method which predicts a non-uniform stress distribution.}
    \label{fig:result}
\end{figure}

Among the devices produced for this study, four of them (D1, D2, D6, and D13) exhibited minimal or negligible wrinkling patterns, making them well-suited candidates for the proposed solution outlined in this study. For these four drums, we quantified stress distributions and validated their accuracy by reconstructing mode shapes using the derived stresses. In figure~\ref{fig:result}, we provide an illustrative example showcasing both the experimental mode shapes and the reconstructed mode shapes for device D1, utilizing the first four distinct mode shapes. It is evident that the obtained stress distribution was able to accurately reconstruct the experimental mode shapes with a high level of accuracy. The results for devices D2, D6, and D13 are presented in supplementary information S9.
 
Moreover, to highlight the fidelity of the proposed method, we present the extracted displacement field and the corresponding non-uniform stress distributions in figure~\ref{fig:result}. It is important to note that Raman spectroscopy, which we utilized for stress measurement, is limited in its ability to detect the non-uniformity of stress within the drum (see supplementary information S10). The predicted stress distribution obtained through Raman spectroscopy appears to be nearly constant and uniform, with high errors across the drum surface. This is due to the fact that the strain values are lower than $0.1\%$, which is the resolution limit of Raman spectroscopy. In the supplementary information S10, a thorough discussion of the Raman spectroscopy measurements and the obtained stress distributions for device D1 is provided. In contrast, the presented methodology is founded upon continuum mechanics which is not dimension-dependent. As a result, the resolution of this method is primarily constrained by the measurement device's capability to discern mode shapes. Consequently, the method's efficacy remains unaffected by the size of the drum or the scale of its strain distributions. This implies that even for small drums with radii on the order of a few micrometers, our methodology can measure strain and stress distributions, regardless of the magnitude of the strains. Hence, this approach remains applicable across a range of scales, encompassing relatively small drums.


To compare the extracted stress distributions and the nominal stress values obtained assuming a uniform tension distribution, we calculated the spatial averages of normal and shear stresses by
\begin{equation}
    \label{eq:avg_stress}
    {\tilde \sigma _{ij}} = \frac{{\int_0^{2\pi } {\int_0^R {{\sigma _{ij}}r{\rm{d}}r{\rm{d}}\theta } } }}{{\pi {R^2}}},
\end{equation}
where \(i, j = \{r, \theta\}\). To measure the robustness of our findings, we systematically adjusted the level of mode shape fitting during the preliminary stage (see equation~(S3)). This variation allowed us to quantitatively determine the associated standard deviation and obtain valuable insights into the sensitivity of our stress distribution analysis. To determine the nominal stress \(\sigma_0\) under the assumption of a uniform tension distribution, we employed the first resonance frequencies, considering them as resonances of an ideal theoretical drum subjected to uniform tension (see equation~(\ref{eq:simpleUnifrom})). Notably, the spatial average of shear stress for all four drums was found to be negligible. However, this was not the case for the values of \(\tilde{\sigma}_{rr}\) and \(\tilde{\sigma}_{\theta\theta}\), which demonstrated appreciable differences. The computed average stress values, obtained through our analysis and assuming uniform stress distribution, are both presented in table~(\ref{table:stress}). 

The differences observed between the average values of $\tilde{\sigma}_{rr}$ and $\tilde{\sigma}_{\theta\theta}$ in table~(\ref{table:stress}) contradict the uniform stress assumption, which posits that $\tilde{\sigma}_{rr} = \tilde{\sigma}_{\theta\theta} = \sigma_0$. Notably, a greater deviation of $\tilde{\sigma}_{rr}$ from $\tilde{\sigma}_{\theta\theta}$ indicates a higher degree of non-uniformity in the stress distribution within the drum. The observed differences between the average values of \(\tilde{\sigma}_{rr}\) and \(\tilde{\sigma}_{\theta\theta}\) raise doubts about the validity of the uniform stress assumption. Specifically, $\sigma_0$ is at least 47\% greater than the mean value of \(\tilde{\sigma}_{rr}\) for each of the drums. This finding suggests that spatially averaging the stress distributions will not yield a uniform stress representation of the overall behavior of the studied drums. Therefore, accounting for the non-uniformities is essential for proper estimation of the built-in stress in ultra-thin membranes.




\newcommand{\ra}[1]{\renewcommand{\arraystretch}{#1}}
\setlength{\tabcolsep}{15pt}

\begin{table}[b]
\centering
\ra{1}
\caption{Spatial average value of non-uniform stress distribution in comparison with nominal stress value assuming uniform stress distribution in the drums}
\begin{tabular}{lccc}
    \toprule
Device  & \(\displaystyle \tilde{\sigma}_{rr}\) (MPa) & \(\displaystyle \tilde{\sigma}_{\theta \theta}\) (MPa) & \(\displaystyle\sigma_0\) (MPa)  \\
\midrule
D1 & \(\displaystyle 3.76 \pm 0.25 \) & \(\displaystyle 2.80 \pm 0.29 \) & 7.05          \\
D2 & \(\displaystyle 0.65 \pm 0.06 \) & \(\displaystyle 0.53 \pm 0.08 \) & 1.36          \\
D6 & \(\displaystyle 0.54 \pm 0.09 \) & \(\displaystyle 0.33 \pm 0.07 \) & 0.94       \\
D13 & \(\displaystyle 1.03 \pm 0.21 \) & \(\displaystyle 0.49 \pm 0.10 \) & 1.52      \\
    \bottomrule
\end{tabular}
\label{table:stress}
\end{table}

Our method's effectiveness is further evident in figure~\ref{fig:result}, where we observe the influence of a free edge on the displacement and stresses of the drum. The microscope image of device D1, as depicted in figure~\ref{fig:experiment}(d), clearly demonstrates that one side of the drum is clamped, while the other side remains unclamped and capable of free movement. Surprisingly, this free edge has influenced the results by exhibiting less radial displacement and consequently lower stresses on the free side. This finding supports the intuition that a free edge allows the drum to mechanically release stresses near the edge.


\section{Discussion}

The proposed methodology addresses a system of nonlinear equations (equation~(\ref{eq:discrete_matrix_form})), under the assumption of a parabolic transverse static displacement field. However, when the drum exhibits initial corrugations or wrinkles that cannot be adequately characterized by such a parabolic displacement field, the system of equations becomes challenging to solve. In such scenarios, the equations need to be solved by providing a suitable initial guess for the transverse displacement field \(W_{i,j}\). To estimate the static transverse displacement field of the drum in the presence of these non-parabolic deformations, corrugations should be experimentally probed. Several techniques are available for measuring these out-of-plane deformations \cite{hiltunen2021ultrastiff,liu2023enhanced}, which can significantly aid with quantifying the tension distribution.


Notably, microscopic images of the drums (see figure S1) do not always reveal signs of transverse bulge or wrinkles, despite their presence. Although the transverse displacement is relatively small compared to the drum's radius ($W_0/R \leq 0.001$), neglecting it in the modal analysis can lead to inaccurate mode shape estimations and ultimately even to failure in reconstructing the experimental mode shapes. Moreover, even minor static transverse asymmetries can affect the expected mode shapes~\cite{Changguo2007, Kukathasan2003}, emphasizing the need for an accurate solution capturing these deviations. This emphasizes the importance of the transverse displacement field when reconstructing the stress distribution.



Owing to inherent experimental uncertainties and noise, there exists a lower bound on the threshold \(e_0\). For device D1, the estimated experimental noise on each mode shape is 3\%, 4\%, 18\%, and 20\%, respectively, leading to a lower average bound for \(e_0\) of 11\%. To enhance the precision of measuring \(e_0\), employing measurement devices with higher spatial resolution and lower noise levels is recommended.



The comprehensive study of drums yielded valuable insights into their stress distributions. These drums experience uniaxial or biaxial loading with different loadings along the two axes, suggesting non-uniform biaxial tension induced in the manufacturing process. This understanding is crucial for optimizing manufacturing processes to achieve uniform stress distribution and flatness in the drums. 

The proposed method is specifically tailored for thin drums, taking into account both stretching and bending energies to derive the governing equations and ensure numerical stability. As a result, two distinct mechanisms govern the mode shapes and resonance frequencies. In cases where stretching dominates the deformation of the drum, the pre-stresses play a significant role in influencing its vibrational behavior. This scenario is particularly relevant for drums with a height-to-radius ratio \(h/R \leq 0.001\). Conversely, as bending deformation becomes more prevalent over stretching, the vibrational behavior of the drum is primarily governed by bending energies, with pre-stresses having a marginal role. In such instances, the accuracy of the proposed method may be compromised, as the mode shapes are predominantly influenced by bending effects rather than stress distributions.

Despite this limitation, in practical applications, the first scenario (\(h/R \leq 0.001\)) is often encountered, rendering the proposed method suitable and reliable for analyzing the vibrational behavior of drums. It is worth noting that as \(h/R\) increases, the bending deformations become more energy costly, resulting in drums with fewer corrugations and wrinkles. Unfortunately, this also leads to reduced sensitivity to transverse loadings and masses. Hence, the choice of \(h/R\) becomes critical in designing circular drums to achieve the desired vibrational characteristics and performance for specific applications.


\section*{Conclusion}

In conclusion, we presented a new methodology for quantitative determination of the tension distribution in ultra-thin 2D material drums based on experimental mode shapes and resonance frequencies. By utilizing a circular plate model, we derived governing equations that capture the static and dynamic deformation of the drums. The proposed approach successfully accounts for both stretching and bending energies, providing an effective solution for analyzing the vibrational behavior of circular drums.

The validation of the method through finite element simulations on known stress distributions demonstrates its accuracy and reliability. We applied the methodology to four fabricated drums and gained valuable insights into their stress distributions. The findings revealed the presence of non-uniform biaxial tension induced during the manufacturing process. Understanding these stress distributions is critical for optimizing the fabrication processes to achieve uniform stress distribution and flatness in the drums.

Additionally, we discussed the limitations related to the assumption of a parabolic transverse static displacement field and the need for an appropriate initial guess for the transverse displacement field in cases where initial corrugations or wrinkles are present. The insights gained from this study can aid in achieving better performance and reliability in 2D drum fabrication and contribute to accurate and robust mechanical characterization of ultra-thin materials.

\section*{Acknowledgement}

This project has received funding from European Union’s Horizon 2020 research and innovation programme under Grant Agreement Nos. 802093 (ERC starting grant ENIGMA), 785219, and 881603 (Graphene Flagship). Katarina Cvetanović and Marko Spasenović acknowledge support by the Serbian Ministry of Science, Technological Development, and Innovations, contract number 451-03-47/2023–01/200026.

\section*{Author declarations}

The authors have no conflicts to disclose.

\section*{Data availability}

The data that support the findings of this study are available from the corresponding author upon reasonable request.

\newgeometry{margin=0cm}
\begin{figure}
  \centering
  \includegraphics[width=\textwidth]{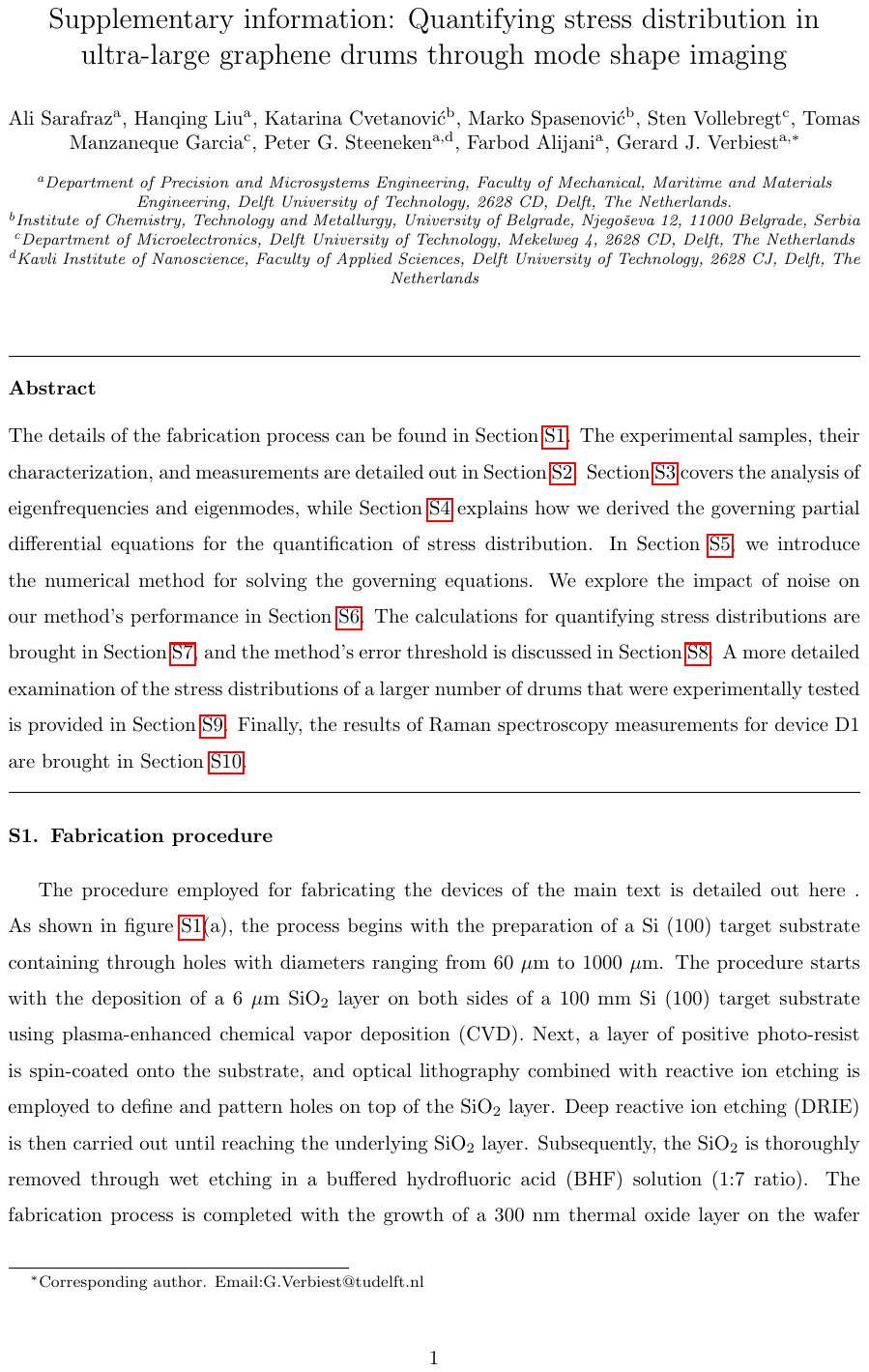}
\end{figure}
\restoregeometry

\clearpage

\afterpage{%
  \newgeometry{margin=0cm} 
  

\section*{Supplementary material}

# Supplementary information: Quantifying stress distribution in ultra-large graphene drums through mode shape imaging

Ali Sarafraz[a], Hanqing Liu[a], Katarina Cvetanović[b], Marko Spasenović[b], Sten Vollebregt[c], Tomas Manzaneque Garcia[c], Peter G. Steeneken[a,d], Farbod Alijani[a], Gerard J. Verbiest[a,*]

[a]*Department of Precision and Microsystems Engineering, Faculty of Mechanical, Maritime and Materials Engineering, Delft University of Technology, 2628 CD, Delft, The Netherlands.*
[b]*Institute of Chemistry, Technology and Metallurgy, University of Belgrade, Njegoševa 12, 11000 Belgrade, Serbia*
[c]*Department of Microelectronics, Delft University of Technology, Mekelweg 4, 2628 CD, Delft, The Netherlands*
[d]*Kavli Institute of Nanoscience, Faculty of Applied Sciences, Delft University of Technology, 2628 CJ, Delft, The Netherlands*

**Abstract**

The details of the fabrication process can be found in Section S1. The experimental samples, their characterization, and measurements are detailed out in Section S2. Section S3 covers the analysis of eigenfrequencies and eigenmodes, while Section S4 explains how we derived the governing partial differential equations for the quantification of stress distribution. In Section S5, we introduce the numerical method for solving the governing equations. We explore the impact of noise on our method's performance in Section S6. The calculations for quantifying stress distributions are brought in Section S7, and the method's error threshold is discussed in Section S8. A more detailed examination of the stress distributions of a larger number of drums that were experimentally tested is provided in Section S9. Finally, the results of Raman spectroscopy measurements for device D1 are brought in Section S10.

**S1. Fabrication procedure**

The procedure employed for fabricating the devices of the main text is detailed out here . As shown in figure S1(a), the process begins with the preparation of a Si (100) target substrate containing through holes with diameters ranging from 60 $\mu$m to 1000 $\mu$m. The procedure starts with the deposition of a 6 $\mu$m SiO$_2$ layer on both sides of a 100 mm Si (100) target substrate using plasma-enhanced chemical vapor deposition (CVD). Next, a layer of positive photo-resist is spin-coated onto the substrate, and optical lithography combined with reactive ion etching is employed to define and pattern holes on top of the SiO$_2$ layer. Deep reactive ion etching (DRIE) is then carried out until reaching the underlying SiO$_2$ layer. Subsequently, the SiO$_2$ is thoroughly removed through wet etching in a buffered hydrofluoric acid (BHF) solution (1:7 ratio). The fabrication process is completed with the growth of a 300 nm thermal oxide layer on the wafer

---

*Corresponding author. Email:G.Verbiest@tudelft.nl

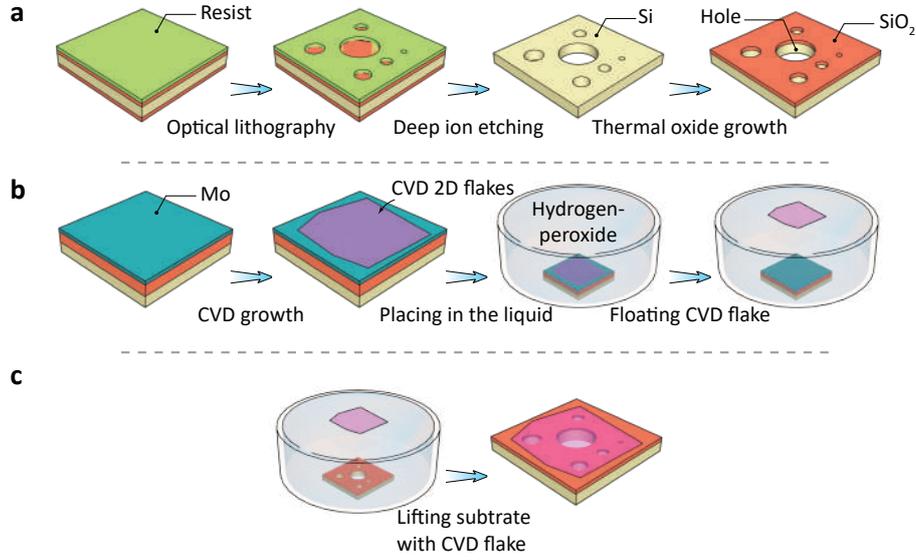

Figure S1: Fabrication and vibration measurement of graphene drums. (a) Fabrication process of SiO$_2$/Si substrate with etched holes. (b) Growth and exfoliation of large-scale CVD graphene flake. (c) Wet transfer method to suspend CVD graphene on substrate.

using a wet oxidation technique.

In the next step, as shown in figure S1(b), multi-layer graphene is deposited using chemical vapor deposition (CVD) on a thin-film Mo catalyst. This process begins with the sputtering of a 50 nm Mo layer onto a 100 mm Si (100) substrate, which is covered by a 600 nm thermal oxide layer [1]. Subsequent CVD is carried out at a substrate temperature of 915 ℃, utilizing H$_2$/CH$_4$ gases at flow rates of 1000/25 sccm, respectively, under a pressure of 25 mbar. The growth process is conducted for 30, 45, 60, and 90 minutes for different samples, after which the wafer is cooled down within an argon ambient environment.

The final stage of the fabrication procedure, as depicted in figure S1(c), involves transferring the CVD-grown graphene from the growth substrate to the target substrate through a wet transfer process. This step begins by immersing the graphene-coated growth substrate in a 30% hydrogen peroxide solution for 25 minutes, facilitating the detachment of the graphene layer, which then floats atop the hydrogen peroxide solution. Subsequent rinsing with deionized (DI) water is performed twice to ensure complete removal of residual hydrogen peroxide. To optimize the transfer, a detergent solution (1 drop of Triton X100 in 150 ml of DI water) is introduced to reduce surface tension. The graphene layer is then picked up using the target substrate, resulting in the formation of suspended drum structures over the fabricated holes. The samples are subsequently dried at room temperature for 25 minutes and left under a glass enclosure for 24 hours.

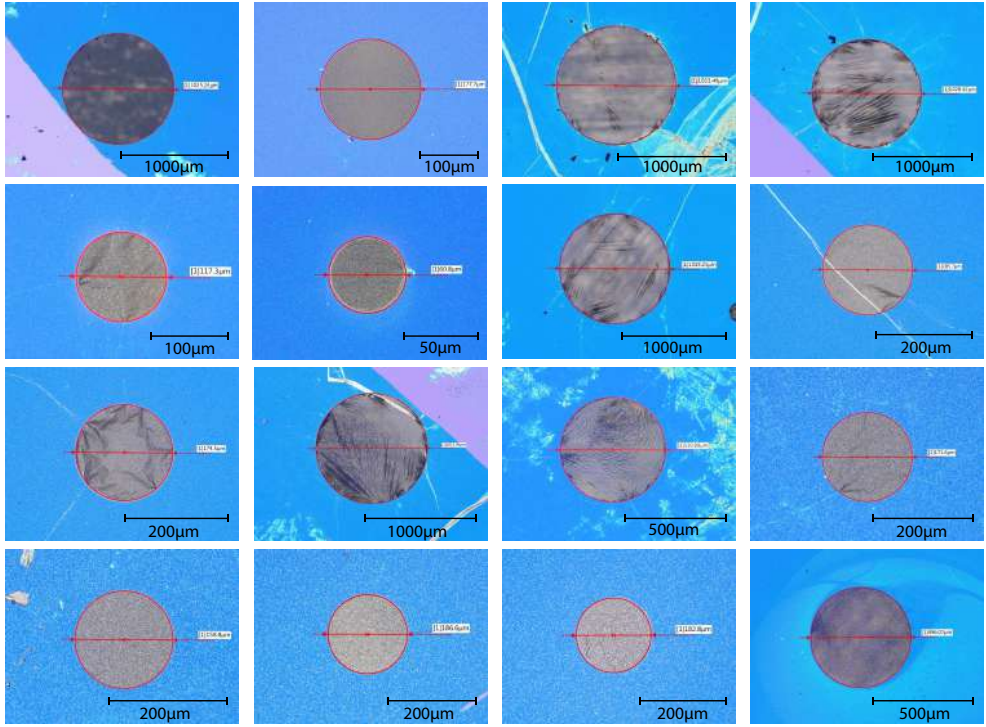

Figure S2: Optical images of the fabricated graphene membranes (devices D1 to D16) on silicon substrate.

## S2. Sample characterization and AFM measurements

Figure S2 shows all graphene drums we fabricated in this work. Using optical microscopy the image of each drum is obtained and then numbered as devices D1 to D16. The diameters of the fabricated drums vary from 60.8 to 1031.5 $\mu$m, as summarized in table S1. Structural defects such as wrinkles that result from the transfer process can extend over the membranes, causing an irregular distribution of graphene sheet tension around its perimeter. Due to the small thickness and high flexibility of graphene, structural defects can be clearly visible on the defected drums.

Here all devices in figures S2 are fabricated onto 6 Si chips, as numbered from chip A to F (see table S1). Assuming that the CVD graphene on each chip is uniform, we use AFM to measure the membrane thickness for all devices. We scan the selected edge area of graphene membranes, as shown in the black frame in figures S3(a). The surface height rises from 16.3 to 23.8 nm, corresponding to a membrane thickness of 7.5 nm (figures S3(b) and S3(c)). As shown in figure S3(d), for chip A, we use the statistics to plot the height histogram for both substrate (black bars) and membrane (red bars). Accordingly, we extract a thickness of 10 nm, corresponding to a surface mass density about $2.267 \times 10^{-5}$ kg/m$^2$. Using this method, we further extract the density of all the other chips as listed in table S1.

3

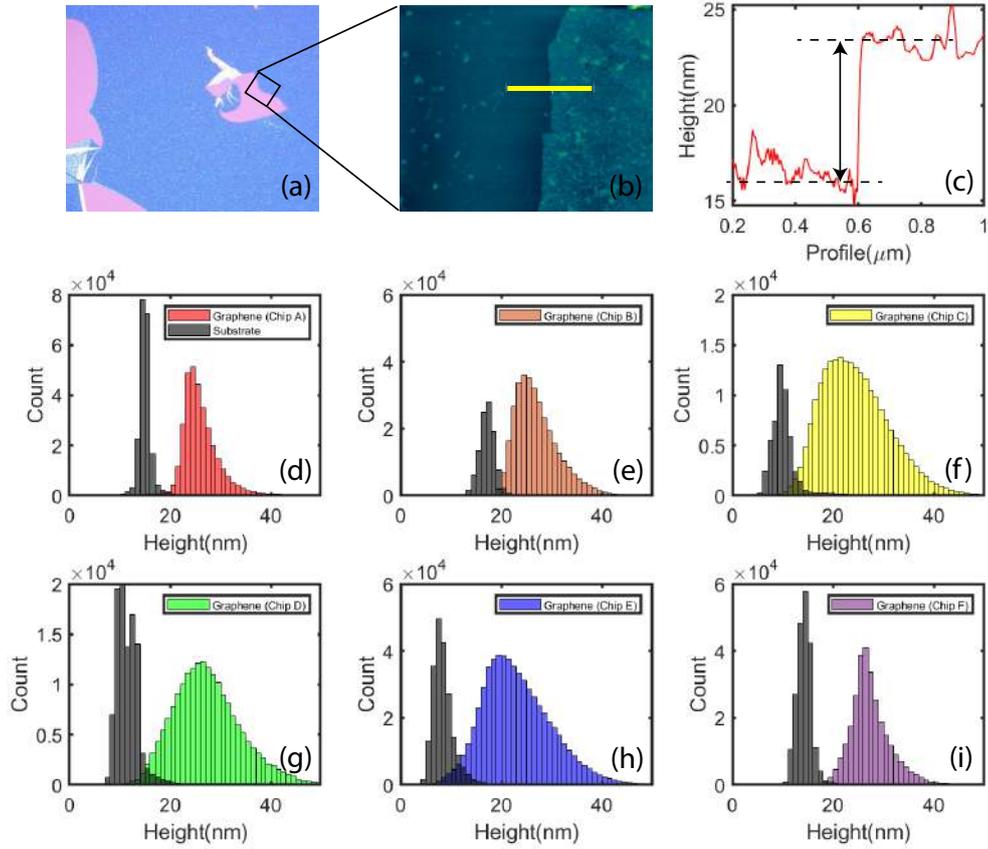

Figure S3: (a) Optical image of the edge of graphene membrane on chip A. (b) AFM image corresponding to the black frame in (a). (c) Height profile along the yellow line shown in (b). (d) Height histogram for both substrate and graphene membrane measured by AFM, allowing us to extract a thickness of around 10 nm. (e)-(i) Height histograms for chip B to F, respectively.

## S3. Measured frequency ratios of experimentally measured graphene drums

Here, we evaluate the ratios of $f_3/f_1$ and $f_4/f_1$ within a particular group of drums. The results of our investigation demonstrate a notable disparity between the experimental observations and the theoretical predictions made under the assumption of uniform pre-tension. Figure S4 shows this analysis for 7 different drums.

## S4. Governing equations

As previously mentioned in the main text, the circular plate model employed in this study is defined by its radius, denoted as $R$, and its thickness, denoted as $h$. The underlying assumption of the model is that the drum is composed of a uniformly distributed and consistent material with density ($\rho$), Young's modulus ($E$), and Poisson's ratio ($\nu$). The governing equations are derived by employing cylindrical coordinates ($r$, $\theta$, $z$), where $r$ denotes the radial coordinate, $\theta$

4

Table S1: Geometrical parameters of suspended graphene samples.

| Chip | Thickness(nm) | Drum No. | Diameter($\mu$m) |
|---|---|---|---|
| A | 10.0 | 1 | 1025.2 |
|   |      | 2 | 177.7 |
|   |      | 3 | 158.0 |
| B | 7.0  | 4 | 1028.4 |
|   |      | 5 | 1031.5 |
|   |      | 6 | 117.3 |
|   |      | 7 | 60.8 |
| C | 11.9 | 8 | 1019.2 |
|   |      | 9 | 185.7 |
| D | 13.8 | 10 | 1011.9 |
|   |      | 11 | 519.8 |
|   |      | 12 | 171.6 |
|   |      | 13 | 158.8 |
| E | 11.9 | 14 | 186.6 |
|   |      | 15 | 182.8 |
| F | 11.9 | 16 | 496.1 |

represents the azimuthal coordinate ($\theta$ direction), and $z$ corresponds to the transverse coordinate. The displacement field of the drums can be mathematically represented as functions of time $t$ by the utilisation of the following equations [2]

$$\begin{aligned} u_r(r,\theta,z,t) &= u(r,\theta,t) - z\frac{\partial w}{\partial r}, \\ u_\theta(r,\theta,z,t) &= v(r,\theta,t) - z\left(\frac{1}{r}\frac{\partial w}{\partial \theta}\right), \\ u_z(r,\theta,z,t) &= w(r,\theta,t), \end{aligned} \quad \text{(S1)}$$

where $(u,v,w)$ represent the radial, tangential, and transverse displacements, respectively. Taking nonlinearities into account, the nonlinear normal and shear strains are determined as

$$\begin{aligned} \varepsilon_{rr} &= \frac{\partial u}{\partial r} + \frac{1}{2}\left(\frac{\partial w}{\partial r}\right)^2 - z\frac{\partial^2 w}{\partial r^2}, \\ \varepsilon_{\theta\theta} &= \frac{u}{r} + \frac{1}{r}\frac{\partial v}{\partial \theta} + \frac{1}{2}\left(\frac{1}{r}\frac{\partial w}{\partial \theta}\right)^2 - \frac{z}{r}\left(\frac{\partial w}{\partial r} + \frac{1}{r}\frac{\partial^2 w}{\partial \theta^2}\right), \\ \varepsilon_{r\theta} &= \frac{1}{2}\left(\frac{1}{r}\frac{\partial u}{\partial \theta} + \frac{\partial v}{\partial r} - \frac{v}{r} + \frac{1}{r}\frac{\partial w}{\partial r}\frac{\partial w}{\partial \theta}\right) - \frac{z}{r}\left(\frac{\partial^2 w}{\partial r\partial \theta} - \frac{1}{r}\frac{\partial w}{\partial \theta}\right). \end{aligned} \quad \text{(S2)}$$

Through constitutive equation and Lamé constants ($\mu = E/(2(1+\nu))$; $\lambda = E\nu/(1-\nu^2)$), the stresses are related to the strain fields as

$$\sigma_{ij} = 2\mu\varepsilon_{ij} + \lambda\delta_{ij}\varepsilon_{kk}, \quad \text{(S3)}$$

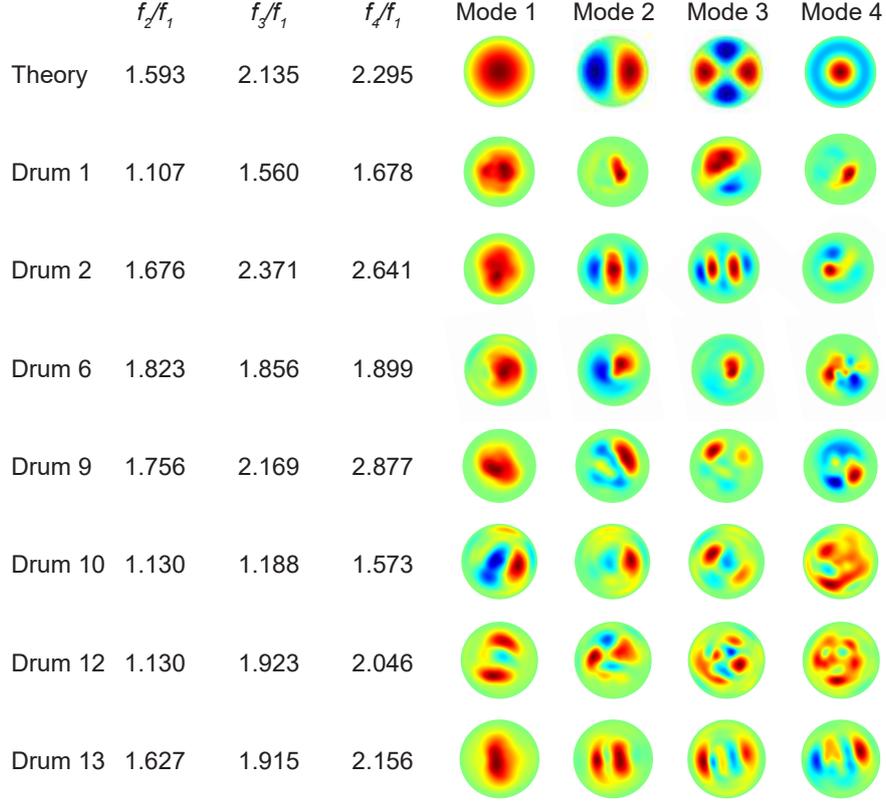

Figure S4: (a) The first, second, third, and fourth mode shapes and the corresponding frequency ratio of seven of the measured devices and their comparison to theoretical estimates based on equation (1) of the main text. All mode shapes are normalized with respect to the maximum value of their displacement, and the colorbar ranges between -1 and 1.

where $\sigma_{ij}$ are the stresses, $\delta_{ij}$ denotes Kronecker delta, and $i, j = \{r, \theta\}$.

The strain energy of the circular plate can be obtained as

$$U = \int_0^{2\pi} \int_0^R \int_{-\frac{h}{2}}^{+\frac{h}{2}} \sigma_{ij}\varepsilon_{ij}\mathrm{d}z\mathrm{d}r\mathrm{d}\theta, \tag{S4}$$

and its kinetic energy neglecting radial (i.e. in-plane) inertia , is given by

$$T = \frac{1}{2}\int_{-\frac{h}{2}}^{+\frac{h}{2}} \int_0^{2\pi} \int_0^R \rho \dot{w}^2 r \mathrm{d}r\mathrm{d}\theta\mathrm{d}z, \tag{S5}$$

where the overdot indicates differentiation with respect to time $t$.

Next, by using the Hamilton's principle

$$\delta \int_0^T (T-U)\,\mathrm{d}t = 0, \tag{S6}$$

in which $\delta$ is the variational operator, the governing equations of motion can be found as follows [2]

$$\delta u : -\frac{1}{r}\left[\frac{\partial}{\partial r}(rN_{rr}) + \frac{\partial N_{r\theta}}{\partial \theta} - N_{\theta\theta}\right] + I_0\frac{\partial^2 u}{\partial t^2} = 0, \tag{S7a}$$

$$\delta v : -\frac{1}{r}\left[\frac{\partial}{\partial r}(rN_{r\theta}) + \frac{\partial N_{\theta\theta}}{\partial \theta} + N_{r\theta}\right] + I_0\frac{\partial^2 v}{\partial t^2} = 0, \tag{S7b}$$

$$\delta w : -\frac{1}{r}\left[\frac{\partial^2}{\partial r^2}(rM_{rr}) - \frac{\partial M_{\theta\theta}}{\partial r} + \frac{1}{r}\frac{\partial^2 M_{\theta\theta}}{\partial \theta^2} + \frac{2}{r}\frac{\partial}{\partial r}\left(r\frac{\partial M_{r\theta}}{\partial \theta}\right)\right]$$
$$+ \frac{\partial}{\partial r}\left(rN_{rr}\frac{\partial w}{\partial r} + N_{r\theta}\frac{\partial w}{\partial \theta}\right) + \frac{1}{r}\frac{\partial}{\partial \theta}\left(N_{\theta\theta}\frac{\partial w}{\partial \theta} + rN_{r\theta}\frac{\partial w}{\partial r}\right) + I_0\ddot{w} = 0, \tag{S7c}$$

where $\delta u$, $\delta v$ and $\delta w$ are the virtual displacements. Furthermore, $I_0 = \rho h$, the resultant forces $N_{ij} = (N_{rr}, N_{\theta\theta}, N_{r\theta})$, and the resultant moments $M_{ij} = (M_{rr}, M_{\theta\theta}, M_{r\theta})$ are defined by

$$\begin{aligned} N_{ij} &= \int_{-\frac{h}{2}}^{+\frac{h}{2}} \sigma_{ij} \mathrm{d}z, \\ M_{ij} &= \int_{-\frac{h}{2}}^{+\frac{h}{2}} \sigma_{ij} z \mathrm{d}z. \end{aligned} \tag{S8}$$

Equation (S7c), which describes the transverse governing equation, can be reformulated in terms of the displacement fields $u$, $v$, and $w$ by utilizing equations (S2), (S3), and (S8) as

$$\begin{aligned}
&\frac{Eh^3}{12(1-\nu^2)}\left[\left(\frac{\partial^4 w}{\partial r^4}\right) + \left(\frac{2}{r}\frac{\partial^3 w}{\partial r^3}\right) - \left(\frac{1}{r^2}\frac{\partial^2 w}{\partial r^2}\right) + \left(\frac{1}{r^3}\frac{\partial w}{\partial r}\right)\right. \\
&\left.+ \left(\frac{1}{r^4}\frac{\partial^4 w}{\partial \theta^4}\right) + \left(\frac{4}{r^4}\frac{\partial^2 w}{\partial \theta^2}\right) - \left(\frac{2}{r^3}\frac{\partial^3 w}{\partial r\partial\theta^2}\right) + \left(\frac{2}{r^2}\frac{\partial^4 w}{\partial r^2\partial\theta^2}\right)\right] \\
&- \frac{Eh}{1-\nu^2}\left[\left(\frac{1}{r^3}u\frac{\partial^2 w}{\partial \theta^2}\right) + \nu\left(\frac{1}{r^2}\frac{\partial u}{\partial r}\frac{\partial^2 w}{\partial \theta^2}\right) + \frac{1-\nu}{2}\left(\frac{1}{r^2}\frac{\partial^2 u}{\partial \theta^2}\frac{\partial w}{\partial r}\right)\right. \\
&+ (1+\nu)\left(\frac{1}{r}\frac{\partial u}{\partial r}\frac{\partial w}{\partial r}\right) + \frac{1+\nu}{2}\left(\frac{1}{r^2}\frac{\partial^2 u}{\partial r\partial\theta}\frac{\partial w}{\partial \theta}\right) + \frac{1+\nu}{2}\left(\frac{1}{r^3}\frac{\partial u}{\partial \theta}\frac{\partial w}{\partial \theta}\right) \\
&\left.+ (1-\nu)\left(\frac{1}{r^2}\frac{\partial u}{\partial \theta}\frac{\partial^2 w}{\partial r\partial\theta}\right) + \left(\frac{\partial^2 u}{\partial r^2}\frac{\partial w}{\partial r}\right) + \nu\left(\frac{u}{r}\frac{\partial^2 w}{\partial r^2}\right) + \left(\frac{\partial u}{\partial r}\frac{\partial^2 w}{\partial r^2}\right)\right] \\
&- \frac{Eh}{1-\nu^2}\left[\left(\frac{1}{r^3}\frac{\partial^2 v}{\partial \theta^2}\frac{\partial w}{\partial \theta}\right) + \left(\frac{1}{r^3}\frac{\partial v}{\partial \theta}\frac{\partial^2 w}{\partial \theta^2}\right) - \frac{1-\nu}{2}\left(\frac{1}{r^2}\frac{\partial v}{\partial r}\frac{\partial w}{\partial \theta}\right)\right. \\
&- \frac{1-\nu}{2}\left(\frac{1}{r^2}\frac{\partial v}{\partial \theta}\frac{\partial w}{\partial r}\right) + \frac{1+\nu}{2}\left(\frac{1}{r}\frac{\partial^2 v}{\partial r\partial\theta}\frac{\partial w}{\partial r}\right) + (1-\nu)\left(\frac{1}{r}\frac{\partial v}{\partial r}\frac{\partial^2 w}{\partial r\partial\theta}\right) \\
&\left.+ \frac{1-\nu}{2}\left(\frac{v}{r^3}\frac{\partial w}{\partial \theta}\right) - (1-\nu)\left(\frac{v}{r^2}\frac{\partial^2 w}{\partial r\partial\theta}\right) - \frac{1-\nu}{2}\left(\frac{1}{r}\frac{\partial^2 v}{\partial r^2}\frac{\partial w}{\partial \theta}\right) + \frac{\nu}{r}\left(\frac{\partial v}{\partial \theta}\frac{\partial^2 w}{\partial r^2}\right)\right] \\
&- \frac{Eh}{1-\nu^2}\left[\frac{3}{2}\left(\frac{1}{r^4}\frac{\partial^2 w}{\partial \theta^2}\left(\frac{\partial w}{\partial \theta}\right)^2\right) - \frac{1}{2}\left(\frac{1}{r^3}\frac{\partial w}{\partial r}\left(\frac{\partial w}{\partial \theta}\right)^2\right) + \frac{1}{2}\left(\frac{1}{r^2}\frac{\partial^2 w}{\partial \theta^2}\left(\frac{\partial w}{\partial r}\right)^2\right)\right. \\
&+ 2\left(\frac{1}{r^2}\frac{\partial w}{\partial \theta}\frac{\partial w}{\partial r}\frac{\partial^2 w}{\partial r\partial\theta}\right) + \frac{1}{2}\left(\frac{1}{r^2}\frac{\partial^2 w}{\partial r^2}\left(\frac{\partial w}{\partial \theta}\right)^2\right) \\
&\left.+ \frac{3}{2}\left(\frac{\partial^2 w}{\partial r^2}\left(\frac{\partial w}{\partial r}\right)^2\right) + \frac{1}{2}\left(\frac{1}{r}\left(\frac{\partial w}{\partial r}\right)^3\right)\right] + I_0\ddot{w} = 0
\end{aligned} \tag{S9}$$

To characterize the mechanical response of a pre-deformed vibrating drum, we decompose the displacement field into a static and a dynamic part as follows [3, 4],

$$\begin{aligned} u(r,\theta,t) &= u_s(r,\theta) \\ v(r,\theta,t) &= v_s(r,\theta) \\ w(r,\theta,t) &= w_s(r,\theta) + w_d(r,\theta,t). \end{aligned} \tag{S10}$$

Here, the subscript $s$ refers to static components, while $d$ represents dynamic deformations. Inserting equation (S10) in equation (S9) we find the governing static and dynamic equations. The

static equation is obtained as

$$
\begin{aligned}
&\frac{Eh^3}{12(1-\nu^2)}\left[\left(\frac{\partial^4 w_s}{\partial r^4}\right) + \left(\frac{2}{r}\frac{\partial^3 w_s}{\partial r^3}\right) - \left(\frac{1}{r^2}\frac{\partial^2 w_s}{\partial r^2}\right) + \left(\frac{1}{r^3}\frac{\partial w_s}{\partial r}\right)\right.\\
&\left.+\left(\frac{1}{r^4}\frac{\partial^4 w_s}{\partial \theta^4}\right) + \left(\frac{4}{r^4}\frac{\partial^2 w_s}{\partial \theta^2}\right) - \left(\frac{2}{r^3}\frac{\partial^3 w_s}{\partial r\partial \theta^2}\right) + \left(\frac{2}{r^2}\frac{\partial^4 w_s}{\partial r^2\partial \theta^2}\right)\right]\\
&-\frac{Eh}{1-\nu^2}\left[\left(\frac{1}{r^3}u_s\frac{\partial^2 w_s}{\partial \theta^2}\right) + \nu\left(\frac{1}{r^2}\frac{\partial u_s}{\partial r}\frac{\partial^2 w_s}{\partial \theta^2}\right) + \frac{1-\nu}{2}\left(\frac{1}{r^2}\frac{\partial^2 u_s}{\partial \theta^2}\frac{\partial w_s}{\partial r}\right)\right.\\
&+(1+\nu)\left(\frac{1}{r}\frac{\partial u_s}{\partial r}\frac{\partial w_s}{\partial r}\right) + \frac{1+\nu}{2}\left(\frac{1}{r^2}\frac{\partial^2 u_s}{\partial r\partial \theta}\frac{\partial w_s}{\partial \theta}\right) + \frac{1+\nu}{2}\left(\frac{1}{r^3}\frac{\partial u_s}{\partial \theta}\frac{\partial w_s}{\partial \theta}\right)\\
&\left.+(1-\nu)\left(\frac{1}{r^2}\frac{\partial u_s}{\partial \theta}\frac{\partial^2 w_s}{\partial r\partial \theta}\right) + \left(\frac{\partial^2 u_s}{\partial r^2}\frac{\partial w_s}{\partial r}\right) + \nu\left(\frac{u_s}{r}\frac{\partial^2 w_s}{\partial r^2}\right) + \left(\frac{\partial u_s}{\partial r}\frac{\partial^2 w_s}{\partial r^2}\right)\right]\\
&-\frac{Eh}{1-\nu^2}\left[\left(\frac{1}{r^3}\frac{\partial^2 v_s}{\partial \theta^2}\frac{\partial w_s}{\partial \theta}\right) + \left(\frac{1}{r^3}\frac{\partial v_s}{\partial \theta}\frac{\partial^2 w_s}{\partial \theta^2}\right) - \frac{1-\nu}{2}\left(\frac{1}{r^2}\frac{\partial v_s}{\partial r}\frac{\partial w_s}{\partial \theta}\right)\right.\\
&-\frac{1-\nu}{2}\left(\frac{1}{r^2}\frac{\partial v_s}{\partial \theta}\frac{\partial w_s}{\partial r}\right) + \frac{1+\nu}{2}\left(\frac{1}{r}\frac{\partial^2 v_s}{\partial r\partial \theta}\frac{\partial w_s}{\partial r}\right) + (1-\nu)\left(\frac{1}{r}\frac{\partial v_s}{\partial r}\frac{\partial^2 w_s}{\partial r\partial \theta}\right)\\
&\left.+\frac{1-\nu}{2}\left(\frac{v_s}{r^3}\frac{\partial w_s}{\partial \theta}\right) - (1-\nu)\left(\frac{v_s}{r^2}\frac{\partial^2 w_s}{\partial r\partial \theta}\right) - \frac{1-\nu}{2}\left(\frac{1}{r}\frac{\partial^2 v_s}{\partial r^2}\frac{\partial w_s}{\partial \theta}\right) + \frac{\nu}{r}\left(\frac{\partial v_s}{\partial \theta}\frac{\partial^2 w_s}{\partial r^2}\right)\right]\\
&-\frac{Eh}{1-\nu^2}\left[\frac{3}{2}\left(\frac{1}{r^4}\frac{\partial^2 w_s}{\partial \theta^2}\left(\frac{\partial w_s}{\partial \theta}\right)^2\right) - \frac{1}{2}\left(\frac{1}{r^3}\frac{\partial w_s}{\partial r}\left(\frac{\partial w_s}{\partial \theta}\right)^2\right) + \frac{1}{2}\left(\frac{1}{r^2}\frac{\partial^2 w_s}{\partial \theta^2}\left(\frac{\partial w_s}{\partial r}\right)^2\right)\right.\\
&+2\left(\frac{1}{r^2}\frac{\partial w_s}{\partial \theta}\frac{\partial w_s}{\partial r}\frac{\partial^2 w_s}{\partial r\partial \theta}\right) + \frac{1}{2}\left(\frac{1}{r^2}\frac{\partial^2 w_s}{\partial r^2}\left(\frac{\partial w_s}{\partial \theta}\right)^2\right)\\
&\left.+\frac{3}{2}\left(\frac{\partial^2 w_s}{\partial r^2}\left(\frac{\partial w_s}{\partial r}\right)^2\right) + \frac{1}{2}\left(\frac{1}{r}\left(\frac{\partial w_s}{\partial r}\right)^3\right)\right] + I_0 \ddot{w}_s = 0
\end{aligned}
\tag{S11}
$$

Subtracting equation (S11) from the result obtained by inserting equation (S10) into equation (S9) yields the nonlinear dynamic governing equation. Linearizing about the static equilibrium, this latter equation about the pre-deformed configuration $(u_s, v_s, w_s)$ yields

$$
\begin{aligned}
&\frac{Eh^3}{12(1-\nu^2)}\left[\left(\frac{\partial^4 w_d}{\partial r^4}\right) + \left(\frac{2}{r}\frac{\partial^3 w_d}{\partial r^3}\right) - \left(\frac{1}{r^2}\frac{\partial^2 w_d}{\partial r^2}\right) + \left(\frac{1}{r^3}\frac{\partial w_d}{\partial r}\right)\right.\\
&\left.+\left(\frac{1}{r^4}\frac{\partial^4 w_d}{\partial \theta^4}\right) + \left(\frac{4}{r^4}\frac{\partial^2 w_d}{\partial \theta^2}\right) - \left(\frac{2}{r^3}\frac{\partial^3 w_d}{\partial r\partial \theta^2}\right) + \left(\frac{2}{r^2}\frac{\partial^4 w_d}{\partial r^2\partial \theta^2}\right)\right]\\
&-\frac{Eh}{1-\nu^2}\left[\left(\frac{1}{r^3}u_s\frac{\partial^2 w_d}{\partial \theta^2}\right) + \nu\left(\frac{1}{r^2}\frac{\partial u_s}{\partial r}\frac{\partial^2 w_d}{\partial \theta^2}\right) + \frac{1-\nu}{2}\left(\frac{1}{r^2}\frac{\partial^2 u_s}{\partial \theta^2}\frac{\partial w_d}{\partial r}\right)\right.\\
&+(1+\nu)\left(\frac{1}{r}\frac{\partial u_s}{\partial r}\frac{\partial w_d}{\partial r}\right) + \frac{1+\nu}{2}\left(\frac{1}{r^2}\frac{\partial^2 u_s}{\partial r\partial \theta}\frac{\partial w_d}{\partial \theta}\right) + \frac{1+\nu}{2}\left(\frac{1}{r^3}\frac{\partial u_s}{\partial \theta}\frac{\partial w_d}{\partial \theta}\right)\\
&\left.+(1-\nu)\left(\frac{1}{r^2}\frac{\partial u_s}{\partial \theta}\frac{\partial^2 w_d}{\partial r\partial \theta}\right) + \left(\frac{\partial^2 u_s}{\partial r^2}\frac{\partial w_d}{\partial r}\right) + \nu\left(\frac{u_s}{r}\frac{\partial^2 w_d}{\partial r^2}\right) + \left(\frac{\partial u_s}{\partial r}\frac{\partial^2 w_d}{\partial r^2}\right)\right]\\
&-\frac{Eh}{1-\nu^2}\left[\left(\frac{1}{r^3}\frac{\partial^2 v_s}{\partial \theta^2}\frac{\partial w_d}{\partial \theta}\right) + \left(\frac{1}{r^3}\frac{\partial v_s}{\partial \theta}\frac{\partial^2 w_d}{\partial \theta^2}\right) - \frac{1-\nu}{2}\left(\frac{1}{r^2}\frac{\partial v_s}{\partial r}\frac{\partial w_d}{\partial \theta}\right)\right.\\
&-\frac{1-\nu}{2}\left(\frac{1}{r^2}\frac{\partial v_s}{\partial \theta}\frac{\partial w_d}{\partial r}\right) + \frac{1+\nu}{2}\left(\frac{1}{r}\frac{\partial^2 v_s}{\partial r\partial \theta}\frac{\partial w_d}{\partial r}\right) + (1-\nu)\left(\frac{1}{r}\frac{\partial v_s}{\partial r}\frac{\partial^2 w_d}{\partial r\partial \theta}\right)\\
&\left.+\frac{1-\nu}{2}\left(\frac{v_s}{r^3}\frac{\partial w_d}{\partial \theta}\right) - (1-\nu)\left(\frac{v_s}{r^2}\frac{\partial^2 w_d}{\partial r\partial \theta}\right) + \frac{1-\nu}{2}\left(\frac{1}{r}\frac{\partial^2 v_s}{\partial r^2}\frac{\partial w_d}{\partial \theta}\right) + \frac{\nu}{r}\left(\frac{\partial v_s}{\partial \theta}\frac{\partial^2 w_d}{\partial r^2}\right)\right]\\
&-\frac{Eh}{1-\nu^2}\left[\frac{3}{2}\frac{1}{r^4}\left(\left(\frac{\partial w_s}{\partial \theta}\right)^2\frac{\partial^2 w_d}{\partial \theta^2} + 2\frac{\partial w_s}{\partial \theta}\frac{\partial^2 w_s}{\partial \theta^2}\frac{\partial w_d}{\partial \theta}\right)\right.\\
&-\frac{1}{2}\left(\frac{1}{r^3}\left(\left(\frac{\partial w_s}{\partial \theta}\right)^2\frac{\partial w_d}{\partial r} + 2\frac{\partial w_s}{\partial \theta}\frac{\partial w_s}{\partial r}\frac{\partial w_d}{\partial \theta}\right)\right)\\
&+\frac{1}{2}\left(\frac{1}{r^2}\left(\left(\frac{\partial w_s}{\partial r}\right)^2\frac{\partial^2 w_d}{\partial \theta^2} + 2\frac{\partial w_s}{\partial r}\frac{\partial^2 w_s}{\partial \theta^2}\frac{\partial w_d}{\partial r}\right)\right)\\
&+2\left(\frac{1}{r^2}\left(\frac{\partial w_s}{\partial \theta}\frac{\partial w_s}{\partial r}\frac{\partial^2 w_d}{\partial r\partial \theta} + \frac{\partial w_s}{\partial \theta}\frac{\partial^2 w_s}{\partial r\partial \theta}\frac{\partial w_d}{\partial r} + \frac{\partial w_s}{\partial r}\frac{\partial^2 w_s}{\partial r\partial \theta}\frac{\partial w_d}{\partial \theta}\right)\right)\\
&+\frac{1}{2}\left(\frac{1}{r^2}\left(\left(\frac{\partial w_s}{\partial \theta}\right)^2\frac{\partial^2 w_d}{\partial r^2} + 2\frac{\partial w_s}{\partial \theta}\frac{\partial^2 w_s}{\partial r^2}\frac{\partial w_d}{\partial \theta}\right)\right)\\
&\left.+\frac{3}{2}\left(\left(\frac{\partial w_s}{\partial r}\right)^2\frac{\partial^2 w_d}{\partial r^2} + 2\frac{\partial w_s}{\partial r}\frac{\partial^2 w_s}{\partial r^2}\frac{\partial w_d}{\partial r}\right) + \frac{1}{2}\left(\frac{1}{r}3\left(\frac{\partial w_s}{\partial r}\right)^2\frac{\partial w_d}{\partial r}\right)\right] + I_0\ddot{w}_d = 0.
\end{aligned}
\tag{S12}
$$

Next, to write equation (S12) in dimensionless form, we use the following non-dimensional parameters

$$
\begin{aligned}
&\bar{w} = \frac{w}{h}, \quad \bar{u} = \frac{R}{h^2}u, \quad \bar{v} = \frac{R}{h^2}v,\\
&\bar{r} = \frac{r}{R}, \quad \bar{t} = \frac{1}{R}\sqrt{\frac{E}{\rho}}t, \quad \Gamma = \frac{h^2}{R^2(1-\nu^2)},
\end{aligned}
\tag{S13}
$$

8

and obtain the following non-dimensional form of (S12)

$$\begin{aligned}
&\frac{\Gamma}{12}\left[\left(\frac{\partial^4\varphi}{\partial\bar{r}^4}\right)+\left(\frac{2}{\bar{r}}\frac{\partial^3\varphi}{\partial\bar{r}^3}\right)-\left(\frac{1}{\bar{r}^2}\frac{\partial^2\varphi}{\partial\bar{r}^2}\right)+\left(\frac{1}{\bar{r}^3}\frac{\partial\varphi}{\partial\bar{r}}\right)\right.\\
&\left.+\left(\frac{1}{\bar{r}^4}\frac{\partial^4\varphi}{\partial\theta^4}\right)+\left(\frac{4}{\bar{r}^4}\frac{\partial^2\varphi}{\partial\theta^2}\right)-\left(\frac{2}{\bar{r}^3}\frac{\partial^3\varphi}{\partial\bar{r}\partial\theta^2}\right)+\left(\frac{2}{\bar{r}^2}\frac{\partial^4\varphi}{\partial\bar{r}^2\partial\theta^2}\right)\right]\\
&-\Gamma\left[\left(\frac{1}{\bar{r}^3}\bar{u}_s\frac{\partial^2\varphi}{\partial\theta^2}\right)+\bar{u}\left(\frac{1}{\bar{r}^2}\frac{\partial\bar{u}_s}{\partial\bar{r}}\frac{\partial^2\varphi}{\partial\theta^2}\right)+\frac{1-\nu}{2}\left(\frac{1}{\bar{r}^2}\frac{\partial^2\bar{u}_s}{\partial\theta^2}\frac{\partial\varphi}{\partial\bar{r}}\right)\right.\\
&+(1+\nu)\left(\frac{1}{\bar{r}}\frac{\partial\bar{u}_s}{\partial\bar{r}}\frac{\partial\varphi}{\partial\bar{r}}\right)+\frac{1+\nu}{2}\left(\frac{1}{\bar{r}^2}\frac{\partial^2\bar{u}_s}{\partial\bar{r}\partial\theta}\frac{\partial\varphi}{\partial\theta}\right)+\frac{1+\nu}{2}\left(\frac{1}{\bar{r}^3}\frac{\partial\bar{u}_s}{\partial\theta}\frac{\partial\varphi}{\partial\theta}\right)\\
&\left.+(1-\nu)\left(\frac{1}{\bar{r}^2}\frac{\partial\bar{u}_s}{\partial\theta}\frac{\partial^2\varphi}{\partial\bar{r}\partial\theta}\right)+\left(\frac{\partial^2\bar{u}_s}{\partial\bar{r}^2}\frac{\partial\varphi}{\partial\bar{r}}\right)+\nu\left(\frac{\bar{u}_s}{\bar{r}}\frac{\partial^2\varphi}{\partial\bar{r}^2}\right)+\left(\frac{\partial\bar{u}_s}{\partial\bar{r}}\frac{\partial^2\varphi}{\partial\bar{r}^2}\right)\right]\\
&-\Gamma\left[\left(\frac{1}{\bar{r}^3}\frac{\partial^2\bar{v}_s}{\partial\theta^2}\frac{\partial\varphi}{\partial\theta}\right)+\left(\frac{1}{\bar{r}^3}\frac{\partial\bar{v}_s}{\partial\theta}\frac{\partial^2\varphi}{\partial\theta^2}\right)-\frac{1-\nu}{2}\left(\frac{1}{\bar{r}^2}\frac{\partial\bar{v}_s}{\partial\bar{r}}\frac{\partial\varphi}{\partial\theta}\right)\right.\\
&-\frac{1-\nu}{2}\left(\frac{1}{\bar{r}^2}\frac{\partial\bar{v}_s}{\partial\theta}\frac{\partial\varphi}{\partial\bar{r}}\right)+\frac{1+\nu}{2}\left(\frac{1}{\bar{r}}\frac{\partial^2\bar{v}_s}{\partial\bar{r}\partial\theta}\frac{\partial\varphi}{\partial\bar{r}}\right)+(1-\nu)\left(\frac{1}{\bar{r}}\frac{\partial\bar{v}_s}{\partial\bar{r}}\frac{\partial^2\varphi}{\partial\bar{r}\partial\theta}\right)\\
&\left.+\frac{1-\nu}{2}\left(\frac{\bar{v}_s}{\bar{r}^3}\frac{\partial\varphi}{\partial\theta}\right)-(1-\nu)\left(\frac{\bar{v}_s}{\bar{r}^2}\frac{\partial^2\varphi}{\partial\bar{r}\partial\theta}\right)+\frac{1-\nu}{2}\left(\frac{1}{\bar{r}}\frac{\partial^2\bar{v}_s}{\partial\bar{r}^2}\frac{\partial\varphi}{\partial\theta}\right)+\frac{\nu}{\bar{r}}\left(\frac{\partial\bar{v}_s}{\partial\theta}\frac{\partial^2\varphi}{\partial\bar{r}^2}\right)\right]\\
&-\Gamma\left[\frac{3}{2}\frac{1}{\bar{r}^4}\left(\left(\frac{\partial\bar{w}_s}{\partial\theta}\right)^2\frac{\partial^2\varphi}{\partial\theta^2}+2\frac{\partial\bar{w}_s}{\partial\theta}\frac{\partial^2\bar{w}_s}{\partial\theta^2}\frac{\partial\varphi}{\partial\theta}\right)\right.\\
&-\frac{1}{2}\left(\frac{1}{\bar{r}^3}\left(\left(\frac{\partial\bar{w}_s}{\partial\theta}\right)^2\frac{\partial\varphi}{\partial\bar{r}}+2\frac{\partial\bar{w}_s}{\partial\theta}\frac{\partial\bar{w}_s}{\partial\bar{r}}\frac{\partial\varphi}{\partial\theta}\right)\right)\\
&+\frac{1}{2}\left(\frac{1}{\bar{r}^2}\left(\left(\frac{\partial\bar{w}_s}{\partial\bar{r}}\right)^2\frac{\partial^2\varphi}{\partial\theta^2}+2\frac{\partial\bar{w}_s}{\partial\bar{r}}\frac{\partial^2\bar{w}_s}{\partial\theta^2}\frac{\partial\varphi}{\partial\bar{r}}\right)\right)\\
&+2\left(\frac{1}{\bar{r}^2}\left(\frac{\partial\bar{w}_s}{\partial\theta}\frac{\partial\bar{w}_s}{\partial\bar{r}}\frac{\partial^2\varphi}{\partial\bar{r}\partial\theta}+\frac{\partial\bar{w}_s}{\partial\theta}\frac{\partial^2\bar{w}_s}{\partial\bar{r}\partial\theta}\frac{\partial\varphi}{\partial\bar{r}}+\frac{\partial\bar{w}_s}{\partial\bar{r}}\frac{\partial^2\bar{w}_s}{\partial\bar{r}\partial\theta}\frac{\partial\varphi}{\partial\theta}\right)\right)\\
&+\frac{1}{2}\left(\frac{1}{\bar{r}^2}\left(\left(\frac{\partial\bar{w}_s}{\partial\theta}\right)^2\frac{\partial^2\varphi}{\partial\bar{r}^2}+2\frac{\partial\bar{w}_s}{\partial\theta}\frac{\partial^2\bar{w}_s}{\partial\bar{r}^2}\frac{\partial\varphi}{\partial\theta}\right)\right)\\
&\left.+\frac{3}{2}\left(\left(\frac{\partial\bar{w}_s}{\partial\bar{r}}\right)^2\frac{\partial^2\varphi}{\partial\bar{r}^2}+2\frac{\partial\bar{w}_s}{\partial\bar{r}}\frac{\partial^2\bar{w}_s}{\partial\bar{r}^2}\frac{\partial\varphi}{\partial\bar{r}}\right)+\frac{1}{2}\left(\frac{1}{\bar{r}}3\left(\frac{\partial\bar{w}_s}{\partial\bar{r}}\right)^2\frac{\partial\varphi}{\partial\bar{r}}\right)\right]=\bar{\omega}^2\varphi.
\end{aligned} \quad (S14)$$

**S5. Discretization of the governing equations**

In this section we discuss how by using vibrational modes and frequencies of the drum, it is possible to find the unknown displacement field $\bar{u}_s$, $\bar{v}_s$, and $\bar{w}_s$ and thus the stress distribution.

Our analysis starts from equation (S14). This equation incorporates the known resonance frequencies $\bar{\omega}$ and mode shapes $\varphi$ as parameters that can be measured from experiments. Naturally to determine the three unknowns $\bar{u}_s$, $\bar{v}_s$, and $\bar{w}_s$ from this equation, one would need to account for minimum three sets of mode shapes ($\varphi_1$, $\varphi_2$, $\varphi_3$) and their corresponding frequencies ($\bar{\omega}_1$, $\bar{\omega}_2$, $\bar{\omega}_3$), and solve the set of nonlinear partial differential equations using a numerical procedure e.g. Finite Element Method (FEM), Finite Difference Method (FDM) or method of Differential Quadratures (DQ).

Here, we make use of the DQ method [5] to solve the equations of motion for the unknown displacement field. The DQ method transforms equation (S14) into a set of nonlinear algebraic equations by approximating partial derivatives of functions at discrete points. We particularly apply polynomial-based Differential Quadrature (PDQ) in the radial direction, while we use Fourier Expansion-based Differential Quadrature (FDQ) in the azimuthal direction. This combination ensures accurate results in both $r$ and $\theta$ directions and handles periodic domains effectively due to FDQ's inherent periodicity.

In order to apply DQ method, one would first need to discretize partial derivatives of given functions. Here, we denote the $a$-th derivative of a function $f(x)$ as $\partial^a f/\partial x^a$. We then express this

derivative as a linear combination of the function values at discrete points along the corresponding coordinate direction, given by [5]

$$\left.\frac{\partial^a f(x)}{\partial x^a}\right|_{x=x_i} = \sum_{j=1}^{n} A_{i,j}^{(a)} f(x_j), \tag{S15}$$

where $n$ is the number of total discrete grid points used in the approximation and $A_{i,j}^{(a)}$ are weighting coefficients. Based on the choice of basis functions for the approximation, different sets of weighting coefficients will be obtained. When PDQ is used, the weighting coefficients of the first derivative are determined as [5]

$$A_{i,j}^{(1)} = \frac{M(x_i)}{(x_i - x_j) M(x_j)} \quad (i, j = 1, 2, ..., n; i \neq j) \tag{S16}$$

where

$$M(x_i) = \prod_{i=1, i \neq j}^{n} (x_i - x_j). \tag{S17}$$

The weighting coefficients of higher-order derivatives can then be obtained through the following recurrence relation

$$A_{i,j}^{(a)} = \begin{cases} a \left[ A_{i,j}^{(a-1)} A_{i,j}^{(1)} - \frac{A_{i,j}^{(a-1)}}{x_i - x_j} \right] & i \neq j \\ -\sum_{j=1}^{n} A_{i,j}^{(a)} & i = j \end{cases}, \tag{S18}$$

where $(i, j = 1, 2, ..., n;\ 2 \leq a \leq n-1)$. However, if FDQ is used, the weighting coefficients of the first derivative are [5]

$$A_{i,j}^{(1)} = \frac{M(x_i)}{(x_i - x_j) M(x_j)} \quad (i, j = 1, 2, ..., n; i \neq j) \tag{S19}$$

where

$$M(x_i) = \prod_{i=1, i \neq j}^{n} \sin\left(\frac{(x_i - x_j)}{2}\right). \tag{S20}$$

The weighting coefficients of higher-order derivatives can be obtained through the following recurrence relation

$$A_{i,j}^{(a)} = \begin{cases} \frac{M(x_i)}{2 \sin\left(\frac{x_i - x_j}{2}\right) M(x_j)} & i \neq j \\ -\sum_{j=1}^{n} A_{i,j}^{(a)} & i = j \end{cases}, \tag{S21}$$

where $(i, j = 1, 2, ..., n;\ 2 \leq a \leq n-1)$. As pointed out earlier, since we have periodicity in the azimutal direction in circular drums, we opt for FDQ in $\theta$ direction and PDQ in the radial. Using these two procedures we then find the following set of discretized derivatives:

$$\begin{aligned} \frac{\partial^a f}{\partial r^a} &= \sum_{k=1}^{N} C_{i,k}^{(a)} F_{k,j}, \\ \frac{\partial^a f}{\partial \theta^a} &= \sum_{k=1}^{M} \bar{C}_{i,k}^{(a)} F_{k,j}, \\ \frac{\partial^{a+b} f}{\partial r^a \partial \theta^b} &= \sum_{k_1=1}^{N} \sum_{k_2=1}^{M} C_{i,k_1}^{(a)} \bar{C}_{j,k_2}^{(b)} F_{k_1, k_2}, \end{aligned} \tag{S22}$$

10

where $f$ is the parameter to be differentiated, $F_{i,j}$ is the value of $f$ at node $(i, j)$, $C_{i,j}^{(a)}$ are the PDQ weight coefficients for domain $\bar{r} \in [0, 1]$, and $\bar{C}_{j,k_2}^{(b)}$ are the FDQ weight coefficients for the domain $\theta \in [0, 2\pi)$.

We note that in estimating stress distributions, the selection of an appropriate mesh distribution is an important factor that can very well affect the accuracy of the obtained displacement field and thus stress distribution. This is particularly important when calculating derivatives at the boundary conditions and ensuring the fulfillment of derivative constraints at the boundaries. Inadequate mesh distribution, characterized by a low number of nodes at the boundaries, may lead to inaccurate stress values near the edges. This likely accounts for the observed diminished accuracy near boundaries, as observed by Waitz et al. [6]. By applying appropriate mesh refinement strategies, we can effectively mitigate potential errors and enhance the overall accuracy of our numerical computations. One effective approach to address this issue is to increase the number of nodes near the edges compared to the main domain. In our DQ analysis, considering the periodicity in the azimuthal direction, we chose a uniform node distribution. However, for the radial direction, where boundary conditions are present, we utilized a Chebyshev-Gauss-Lobatto distribution [7]:

$$\bar{r}_i = \frac{1}{2}\left[1 - \cos\left(\frac{i-1}{N-1}\pi\right)\right]. \quad i = 1,\ 2,\ \ldots,\ N \tag{S23}$$

This distribution is defined as a non-uniform spacing of nodes which concentrates more nodes near the boundaries and fewer nodes in the central region. The Chebyshev-Gauss-Lobatto distribution ensures better resolution near the boundaries, where accurate approximation of derivatives is important, while maintaining an appropriate node spacing throughout the domain.

Applying boundary conditions is another important consideration when solving the equations using the DQ method. In the $\theta$-direction, the periodicity is automatically satisfied due to the FDQ formulation. However, in the radial direction ($\bar{r}$-direction), since the DQ method discretizes the strong form of the 4th-order partial differential equation, we need to specify two boundary conditions for each of the edges corresponding to $\bar{r} = 0$ and $\bar{r} = 1$. This differs from FEM, which solves the weak form of the governing equations. The boundary condition for $\bar{r} = 1$ is a clamped condition which can be formulated as

$$\begin{aligned}&\Phi_{N,j} = 0, \\ &\left.\frac{\partial \varphi}{\partial \bar{r}}\right|_{\bar{r}=1} = \sum_{k=1}^{N} C_{N,k}^{(1)} \Phi_{k,j} = 0.\end{aligned} \tag{S24}$$

In contrast to rectangular geometry, which features well-defined boundaries, circular geometry lacks a tangible edge at the center. Analytical formulations address this challenge by omitting terms that approach infinity at the center of the circular membrane/plate. However, determining the boundary condition (regularity condition) at the center in a numerical approach is not straightforward, and various methods exist to tackle this issue [8]. Among different approaches, the most

promising regularity boundary condition is defined for two different types of mode shapes. In this approach mode shapes that exhibit nodal lines are subject to the regularity condition expressed as follows [8]

$$\left.\frac{\partial \varphi}{\partial \bar{r}}\right|_{\bar{r}=0} = \sum_{k=1}^{N} C_{1,k}^{(1)} \Phi_{k,j} = 0,$$
$$\left.\frac{\partial^3 \varphi}{\partial \bar{r}^3}\right|_{\bar{r}=0} = \sum_{k=1}^{N} C_{1,k}^{(3)} \Phi_{k,j} = 0.$$
(S25)

While for mode shapes without nodal lines, the regularity condition can be expressed as [8]

$$\Phi_{1,j} = 0,$$
$$\left.\frac{\partial^2 \varphi}{\partial \bar{r}^2}\right|_{\bar{r}=0} = \sum_{k=1}^{N} C_{1,k}^{(2)} \Phi_{k,j} = 0.$$
(S26)

As a result by solving equations (S25) and equations (S24) simultaneously, we find

$$\Phi_{1,j} = \frac{1}{G} \sum_{k=3}^{N-2} G_{1,k} \Phi_{k,j},$$
$$\Phi_{2,j} = \frac{1}{G} \sum_{k=3}^{N-2} G_{2,k} \Phi_{k,j},$$
$$\Phi_{N-1,j} = \frac{1}{G} \sum_{k=3}^{N-2} G_{N-1,k} \Phi_{k,j},$$
(S27)

where

$$G = \left(C_{1,N-1}^{(1)} C_{1,2}^{(3)} - C_{1,2}^{(1)} C_{1,N-1}^{(3)}\right) C_{N,1}^{(1)} + \left(C_{1,1}^{(1)} C_{1,N-1}^{(3)} - C_{1,N-1}^{(1)} C_{1,1}^{(3)}\right) C_{N,2}^{(1)} + \left(C_{1,2}^{(1)} C_{1,1}^{(3)} - C_{1,1}^{(1)} C_{1,2}^{(3)}\right) C_{N,N-1}^{(1)},$$
$$G_{1,k} = \left(C_{1,2}^{(3)} C_{N,N-1}^{(1)} - C_{1,N-1}^{(3)} C_{N,2}^{(1)}\right) C_{1,j}^{(1)} + \left(C_{1,N-1}^{(1)} C_{N,2}^{(1)} - C_{1,2}^{(1)} C_{N,N-1}^{(1)}\right) C_{1,j}^{(3)} + \left(C_{1,2}^{(1)} C_{1,N-1}^{(3)} - C_{1,N-1}^{(1)} C_{1,2}^{(3)}\right) C_{N,j}^{(1)},$$
$$G_{2,k} = \left(C_{1,N-1}^{(3)} C_{N,1}^{(1)} - C_{1,1}^{(3)} C_{N,N-1}^{(1)}\right) C_{1,j}^{(1)} + \left(C_{1,1}^{(1)} C_{N,N-1}^{(1)} - C_{1,N-1}^{(1)} C_{N,1}^{(1)}\right) C_{1,j}^{(3)} + \left(C_{1,N-1}^{(1)} C_{1,1}^{(3)} - C_{1,1}^{(1)} C_{1,N-1}^{(3)}\right) C_{N,j}^{(1)},$$
$$G_{N-1,k} = \left(C_{1,1}^{(3)} C_{N,2}^{(1)} - C_{1,2}^{(3)} C_{N,1}^{(1)}\right) C_{1,j}^{(1)} + \left(C_{1,2}^{(1)} C_{N,1}^{(1)} - C_{1,1}^{(1)} C_{N,2}^{(1)}\right) C_{1,j}^{(3)} + \left(C_{1,1}^{(1)} C_{1,2}^{(3)} - C_{1,2}^{(1)} C_{1,1}^{(3)}\right) C_{N,j}^{(1)}.$$
(S28)

And by solving equations (S26), and equations (S24) we obtain

$$\Phi_{1,j} = 0$$
$$\Phi_{2,j} = \frac{1}{H} \sum_{k=3}^{N-2} H_{2,k} \Phi_{k,j},$$
$$\Phi_{N-1,j} = \frac{1}{H} \sum_{k=3}^{N-2} H_{N-1,k} \Phi_{k,j},$$
(S29)

where

$$H = C_{N,2}^{(1)} C_{1,N-1}^{(2)} - C_{1,2}^{(2)} C_{N,N-1}^{(1)},$$
$$H_{2,k} = C_{N,N-1}^{(1)} C_{1,k}^{(2)} - C_{1,N-1}^{(2)} C_{N,k}^{(1)},$$
$$H_{N-1,k} = C_{1,2}^{(2)} C_{N,k}^{(1)} - C_{N,2}^{(1)} C_{1,k}^{(2)}.$$
(S30)

Thus, for the derivatives of $\bar{w}$, equation (S22) will be modified as

$$\frac{\partial^a \varphi}{\partial r^a} = C_{i,1}^{(a)} \Phi_{1,j} + C_{i,2}^{(a)} \Phi_{2,j} + C_{i,N-1}^{(a)} \Phi_{N-1,j} + \sum_{k=3}^{N-2} C_{i,k}^{(a)} \Phi_{k,j},$$
$$\frac{\partial^a \varphi}{\partial \theta^a} = \sum_{k=1}^{M} \bar{C}_{i,k}^{(a)} \Phi_{k,j},$$
$$\frac{\partial^{a+b} \varphi}{\partial r^a \partial \theta^b} = \sum_{k_2=1}^{M} C_{i,1}^{(a)} \bar{C}_{j,k_2}^{(b)} \Phi_{1,k_2} + \sum_{k_2=1}^{M} C_{i,2}^{(a)} \bar{C}_{j,k_2}^{(b)} \Phi_{2,k_2} + \sum_{k_2=1}^{M} C_{i,N-1}^{(a)} \bar{C}_{j,k_2}^{(b)} \Phi_{N-1,k_2}$$
$$+ \sum_{k_1=3}^{N-2} \sum_{k_2=1}^{M} C_{i,k_1}^{(a)} \bar{C}_{j,k_2}^{(b)} \Phi_{k_1,k_2},$$
(S31)

12

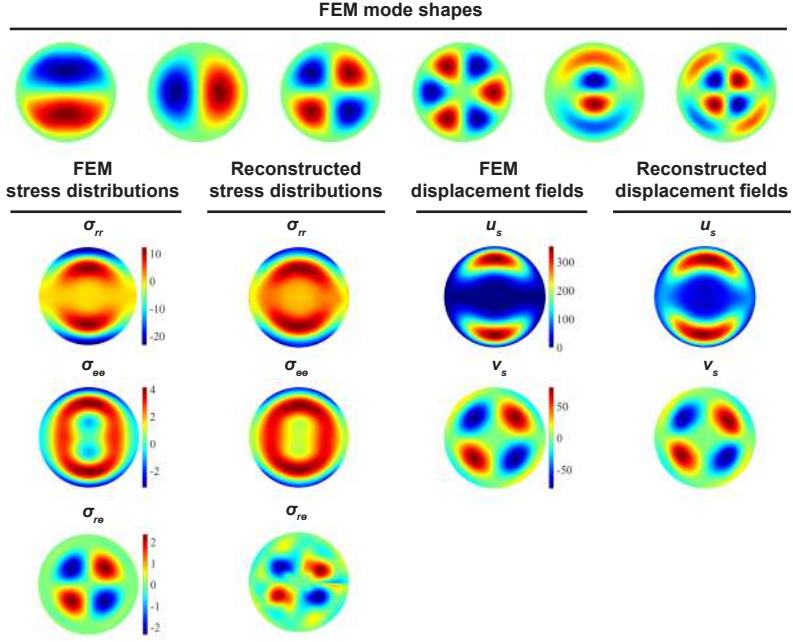

Figure S5: Reconstruction of stress distribution and displacement field for an artificial FEM case. Within the FEM software, displacement fields are generated, and the corresponding stress distributions and mode shapes are computed. Equation (S32) is then used to reconstruct displacement fields from FEM mode shapes. The displacement and stress values are shown in the non-dimensional form.

where, based on the type of mode shape used, $\Phi_{1,j}$, $\Phi_{2,j}$, and $\Phi_{N-1,j}$ should be obtained from (S27) or equations (S29). As can be seen, the modified derivatives are no longer applied to the whole domain, but only to the interior domain without the two adjacent rows of nodes to the edges.

Finally, the discretized version of equation (S14) can be written in a matrix-form as

$$\begin{aligned}\sum_{k_u}\left[\left(\mathbf{D}_{u\varphi}^{k_u}\Phi_{i,j}\right)\mathbf{D}_U^{k_u}\right]U_{i,j}+\sum_{k_v}\left[\left(\mathbf{D}_{v\varphi}^{k_v}\Phi_{i,j}\right)\mathbf{D}_V^{k_v}\right]V_{i,j}\\+\sum_{k_w}\left[\mathbf{D}_{w\varphi}^{k_w}\Phi_{i,j}\right]\left[\left(\bar{\mathbf{D}}_{\mathbf{W}}^{k_w}W_{i,j}\right)\cdot\left(\bar{\bar{\mathbf{D}}}_{\mathbf{W}}^{k_w}W_{i,j}\right)\right]=\left(\bar{\omega}^2 I-\mathbf{D_W}\right)\Phi_{i,j},\end{aligned} \quad (S32)$$

where $U_{i,j}$, $V_{i,j}$, and $W_{i,j}$ denote the static deformation values at each nodes and are the vectors of unknowns. $\Phi_{i,j}$ and $\bar{\omega}$ denote normalized mode shape values at each node and the corresponding resonance frequency, respectively, and are known parameter. $\mathbf{D}_{u\varphi}^{k_u}$, $\mathbf{D}_{v\varphi}^{k_v}$, $\mathbf{D}_{w\varphi}^{k_w}$, $\mathbf{D}_U^{k_u}$, $\mathbf{D}_V^{k_v}$, $\mathbf{D_W}$, $\bar{\mathbf{D}}_{\mathbf{W}}^{k_w}$, and $\bar{\bar{\mathbf{D}}}_{\mathbf{W}}^{k_w}$ are matrices which are function of DQ weighting coefficients.

## S6. Robustness of the DQ solution

We note that equation (S32) invovles nonlinear terms solely related to transverse static displacement field. If the drum is flat ($W_{i,j}=0$), the equation will be considerably simplified (one would solely need to solve for $U_{i,j}$, and $V_{i,j}$). Consequently, instead of having $3\times M\times N$ unknowns (where $M$ represents the number of nodes in the $\theta$ direction and $N$ signifies the number of nodes

in the $\bar{r}$ direction), we will only have $2 \times M \times N$ unknowns. This suggests that two mode shapes and their corresponding resonance frequencies will be sufficient for extracting the unknown displacements. However, it is important to note that the nodes on the clamped edge have a value of zero, which results in $\Phi_{i,j} = 0$ on the edge, and consequently, we have the trivial $0 = 0$ equation on the edge nodes, indicating that we still require at least three mode shapes and frequencies to have more equations than unknowns and be able to extract displacement fields using a least squares procedure.

To assess the validity of our methodology and equations, we initiated a verification procedure by artificially generating a complex radial displacement field represented as

$$u(r,\theta) = 5 \times 10^{-6} r J_6 \left(9.93611 \frac{r}{R}\right) (1 + \sin(2\theta)). \tag{S33}$$

This artificial displacement was then introduced into a FEM software package with the constraint $w = 0$. The resulting equilibrium displacement field $v$ was subsequently obtained. Employing the values of $u$, $v$, and $w$, we computed the corresponding stress distributions. Following this, a modal analysis was conducted to ascertain the resultant mode shapes.

Next, we employed the FEM mode shapes and their corresponding resonance frequencies as inputs in our equation (S32) to reconstruct the displacement fields and stress distributions. The results of this reconstruction using six different mode shapes are shown in figure (S5). Notably, the reconstructed displacement fields exhibit high accuracy, closely resembling the FEM displacement field. Similarly, the calculation of normal stresses demonstrates decent precision. However, it is important to recognize that shear stress accuracy is relatively lower in comparison. This limitation stems from shear stress's substantially smaller amplitude as compared to normal stresses, as well as its comparable amplitude to numerical error created during the differentiation process.

Another crucial aspect to consider is the robustness of our method in the presence of noise and inaccuracies in the measured mode shapes. To evaluate this, we introduced random noise to the mode shapes obtained from the previous analysis. Specifically, we added random two-dimensional Gaussian noise with a maximum amplitude of less than 0.05 times the local deflection to the original mode shapes shown in figure (S5).

Subsequently, we employed a fitting procedure for the noisy mode shapes, which is detailed in the main text and the following section. This fitting procedure employs a finite set of circular plate mode shapes as

$$\begin{aligned}\varphi(r,\theta) = &\sum_{m=0}^{M_f} \sum_{n=0}^{N_f} \left[A_n J_n(\lambda_{m,n} \tfrac{r}{R}) + B_n I_n(\lambda_{m,n} \tfrac{r}{R})\right] \cos(n\theta) \\ &+ \sum_{m=0}^{M_f} \sum_{n=1}^{N_f} \left[A_n^* J_n(\lambda_{m,n} \tfrac{r}{R}) + B_n^* I_n(\lambda_{m,n} \tfrac{r}{R})\right] \sin(n\theta),\end{aligned} \tag{S34}$$

to approximate and enhance the quality of the noisy mode shapes. Here, $M_f$, and $N_f$ are the number of radial and azimuthal coordinates chosen for fitting. $A_n$, $A_n^*$, $B_n$, and $B_n^*$ are the fitting

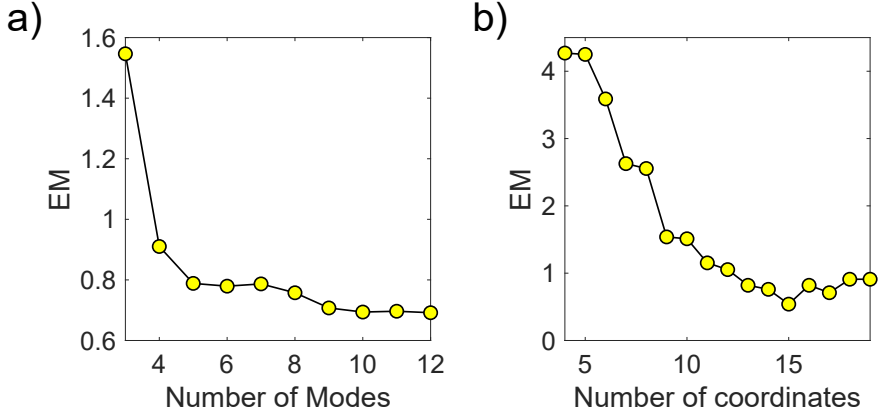

Figure S6: DQ solution's robustness in the presence of noise on measured mode shapes. a) The error metric (S36) of the proposed solution as a function of the number of experimental mode shapes employed in the solution procedure. As demonstrated, adding more mode shapes reduces the error of the solution by averaging the noise in the measured mode shapes; however, after five modes, adding more mode shapes does not significantly improve accuracy. (b) The error metric (S36) of the proposed solution as a function of the number of fitting coordinates employed to fit the mode shapes (see equation (S34)). Fewer coordinates result in greater smoothing, whereas a larger number of coordinates follows all local curvatures, which may be artefacts of noise or measurement errors.

parameters which can be found through the fitting procedure. Here, $J_n$, and $I_n$ denote Bessel function of the first kind of $n$-th order, and modified Bessel function of first kind of the $n$-th order, respectively. $\lambda_{m,n}$ represents the corresponding eigenfrequency of the plate which can be found from the following equation [2]

$$\frac{J_{n+1}(\lambda_{m,n})}{J_n(\lambda_{m,n})} + \frac{I_{n+1}(\lambda_{m,n})}{I_n(\lambda_{m,n})} = 0. \tag{S35}$$

We evaluated the method's accuracy based on the implications of varying the number of experimental mode shapes employed in the fitting procedure (See Figure S6a) and the number of plate modes (coordinates) utilized in the fitting process (See Figure S6b). This allowed us to conduct a parametric study to ascertain whether a good fit, encompassing both the true mode shape and the spurious displacements arising from noise, or a smoother fit with a lower number of coordinates (and accordingly lower $R^2$ value), proves to be more suitable. In order to evaluate the accuracy of our method, we introduce an error metric as

$$EM = \left[ \frac{\int_0^{2\pi} \int_0^R \left(\sigma_{rr}^{predict} - \sigma_{rr}^{real}\right)^2 r dr d\theta}{\int_0^{2\pi} \int_0^R (\sigma_{rr}^{real})^2 r dr d\theta} \right]^{1/2} \tag{S36}$$

where $\sigma_{rr}^{predict}$ denotes the predicted normal stress obtained using equation (S29), and $\sigma_{rr}^{real}$ is the normal stress that is artificially created at the beginning of the procedure. The results of our analysis indicated that we require a fitting procedure with a high degree of accuracy, but also a certain level of smoothing to mitigate the impact of noise (figure (S6)).

15

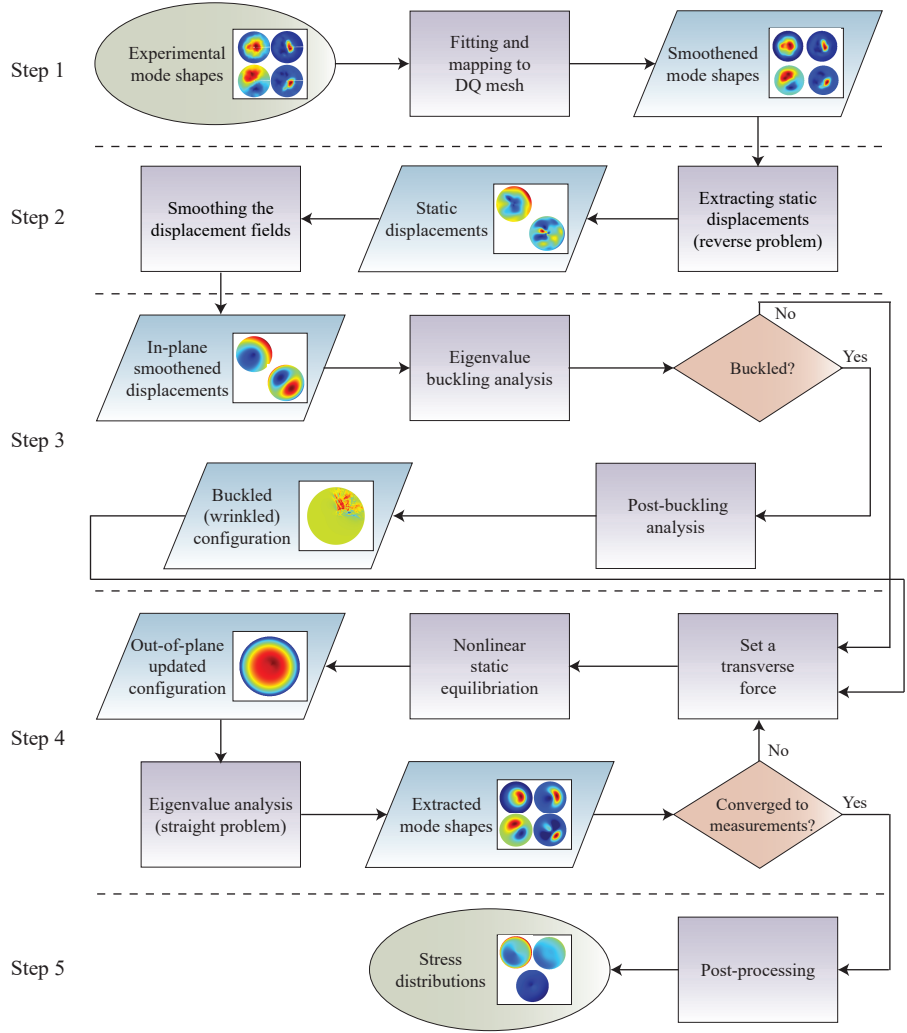

Figure S7: The flowchart for deriving stress distributions from experimental mode shapes and resonance frequencies.

### S7. Extracting the stress distributions

To determine the static displacement fields ($U_{i,j}$, $V_{i,j}$, and $W_{i,j}$) in suspended drums, we expand the methodology explained earlier on numerical simulations to experimental measurements so as to extract the nonuniform stress distributions. The procedure, illustrated in Figure S7, involves several steps to ensure accurate and reliable results. A minimum of three sets of mode shapes and their corresponding resonance frequencies are required as inputs. These mode shapes capture the vibrational behavior of the drums and serve as the basis for reconstructing the displacement fields and stress distributions. It is important to note that our method is applicable only to drums with small deviations from a parabolically-bulged configuration.

To prepare the experimental data for numerical analysis, we address several challenges associated with the mode shapes. These include noise, discreteness, non-zero values on boundaries,

16

and the need for a smooth and continuous representation of the mode shapes. To tackle these challenges, the mode shapes are transformed to polar coordinates and fitted with a plane to remove edge effects and establish zero value at the boundaries. The resulting mode shapes are then smoothed and transferred to the DQ mesh, ensuring that the fitted shapes satisfy the boundary conditions necessary for accurate numerical analysis. Thus, we chose to use plate mode shapes as the basis for fitting the processed experimental mode shapes. The fitting equation used is equation (S34) defined in the last section.

Having obtained the fitted mode shapes, we proceed to solve the reverse problem i.e. finding displacement field from resonance frequencies and mode shapes, by assuming a parabolic transverse deflection profile for the out-of-plane displacement field using equation (S32). The static displacement fields are determined by solving the nonlinear algebraic equations (S32). A least squares procedure is employed, to ensure accurate reconstruction of the in-plane displacements and a first approximation of the out-of-plane deflection. For the sake of numerical robustness, we applied a smoothing operation to the obtained radial displacement field $u$ using a Savitzky-Golay Finite Impulse Response (FIR) filter. Specifically, when the magnitudes of the in-plane displacement field $v$ were comparable to those of $u$, a Savitzky-Golay FIR filter was also employed for smoothing $v$. Conversely, in cases where the values of $v$ were significantly smaller than those of $u$, we exclusively utilized a FEM solver to determine the equilibrium state of $v$. This decision arises from the acknowledgment that when $v$ is significantly smaller than $u$, it becomes more vulnerable to numerical errors, as its values may either approach or be in the vicinity of the numerical error order.

An important consideration is the validation of the initial approximation. To assess its adequacy, an eigenvalue buckling analysis is performed. If the in-plane displacement field leads to a stable drum without buckling, the initial approximation is deemed satisfactory. However, when non-uniform in-plane displacements result in instability and buckling, a post-buckling analysis is conducted to obtain a more realistic deformation pattern.

In the fourth step of our analysis, we evaluate the adequacy of the in-plane displacement fields and the deformed configuration obtained thus far. However, it is often the case that these results do not meet the criterion expressed in equation (5) of the main text as

$$e = \frac{1}{\bar{N}} \sum_{n=1}^{\bar{N}} \left[ \frac{\int_0^{2\pi} \int_0^1 (\bar{\varphi}_n - \varphi_n)^2 r \mathrm{d}r \mathrm{d}\theta}{\int_0^{2\pi} \int_0^1 (\varphi_n)^2 r \mathrm{d}r \mathrm{d}\theta} \right]^{1/2} \leq e_0, \tag{S37}$$

To address this, we introduce a perturbation to further refine our solution. We begin by applying a zero-pressure perturbation and perform a nonlinear static equilibrium analysis using an FEM software package (In our case Ansys was used). This analysis yields an updated out-of-plane configuration in conjunction with the smoothened in-plane displacement fields.

To validate the convergence of our solution, the updated configuration and in-plane displace-

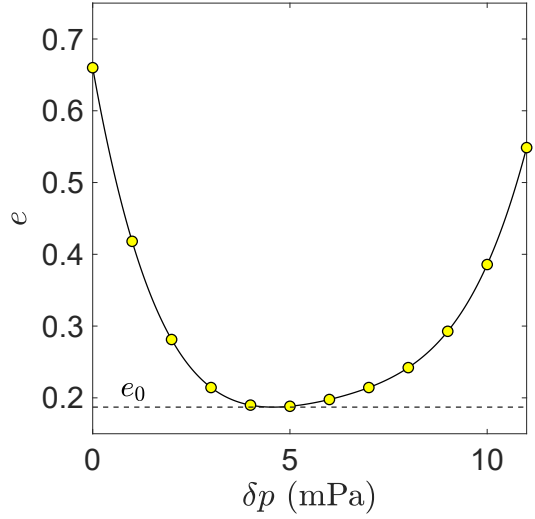

Figure S8: Error criterion $e$ as a function of perturbation uniform transverse load $\delta p$.

ment fields are employed in an eigenvalue modal analysis. By comparing the resulting mode shapes to the experimental mode shapes using the criteria outlined in equation (S37), we then assess the convergence of our solution. However, if the solution fails to converge, we iterate the process by adjusting the perturbation (perturbing out-of-plane force) until convergence is achieved.

This iterative procedure ensures that our analysis reaches a stable and accurate solution, capturing the intricate interplay between in-plane and out-of-plane displacements and providing reliable stress distributions in suspended drums.

In the final step of our analysis, after obtaining convergence, we utilize the static displacement fields (in-plane and transverse) to calculate the strain and stress fields within the drum. This calculation is performed using equations (S2) and (S3) for the mid-plane of the plate. The resulting stress fields serve as a reasonable approximation for the actual experimental stress distributions, as they successfully reproduce the observed mode shapes.

### S8. Finding the minimum error threshold

The error criterion, as presented in equation (S37), is formulated based on the spatial disparity between the experimental mode shapes and the reconstructed mode shapes. It is clear that the mode shapes undergo adjustments through the tuning of a transverse perturbation pressure during the post-buckling analysis S7.

To ascertain an appropriate value for $e_0$, we initiate an iterative procedure without applying any perturbation pressure and gradually increase it. Initially, the observed trend reveals a decreasing error metric $e$ until reaching its minimum value $e_0$ at a perturbation pressure of $\delta p_m$. Beyond this point, further increments in the perturbation pressure result in an increase in error. Consequently,

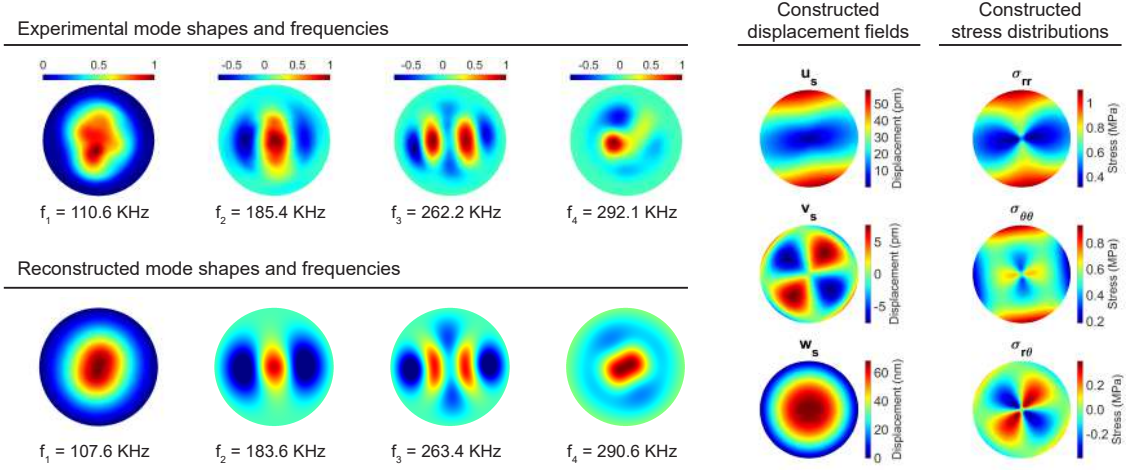

Figure S9: Estimation of stress distributions from experimental mode shapes for device D2.

the minimal achievable error for each set of experiments corresponds to $e_0$, which may vary among different drums. Figure S8 demonstrates this analysis conducted for device D1. To establish the minimum achievable error threshold using our method, this analysis must be performed for each drum under investigation. The plot clearly indicates the error threshold attained in this case as $e_0 \simeq 0.19$ at $\delta p_m \simeq 4.6 \, \text{mPa}$.

The determination of $e_0$ holds importance in ensuring the accuracy and reliability of our method for analyzing the vibrational behavior of circular drums. The specific value of $e_0$ for each drum is significant as it serves as a reference point for assessing the quality of the reconstructed mode shapes and stress distributions. A lower $e_0$ signifies a higher level of fidelity in our predictions, thereby greater confidence in the obtained results.

## S9. Stress distribution in the fabricated devices

As shown in the main text, 4 drums were successfully fitted using the proposed procedure. Here, you can see the experimental mode shapes, reconstructed mode shapes, and also the resulting stress distributions in figures(S9-S11).

## S10. Raman spectroscopy measurements and their limitations

The Raman spectra of graphene is related to the strain loaded on the membrane [9], for example, the slope $\partial \omega / \partial \epsilon$ of 2D peak is about -64 cm$^{-1}$/% [9]. Here, we measure the Raman spectrum of the device D1 using 514 nm green laser at room temperature, as shown in figure S12 (left). Accordingly, we extract the stress distribution of device D1 as plotted in figure S12 (right), which shows an extremely large error bar. This is because of the limited resolution of strain-dependent

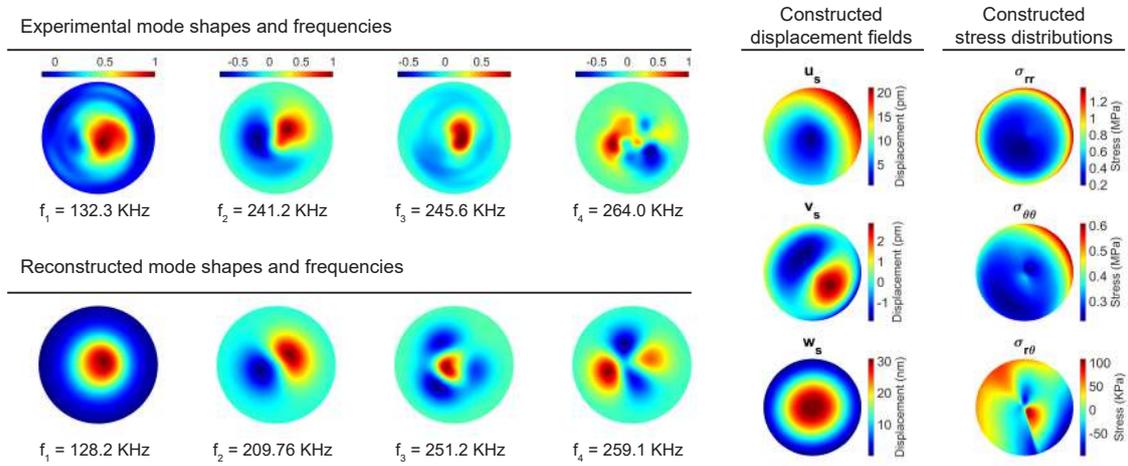

Figure S10: Estimation of stress distributions from experimental mode shapes for device D6.

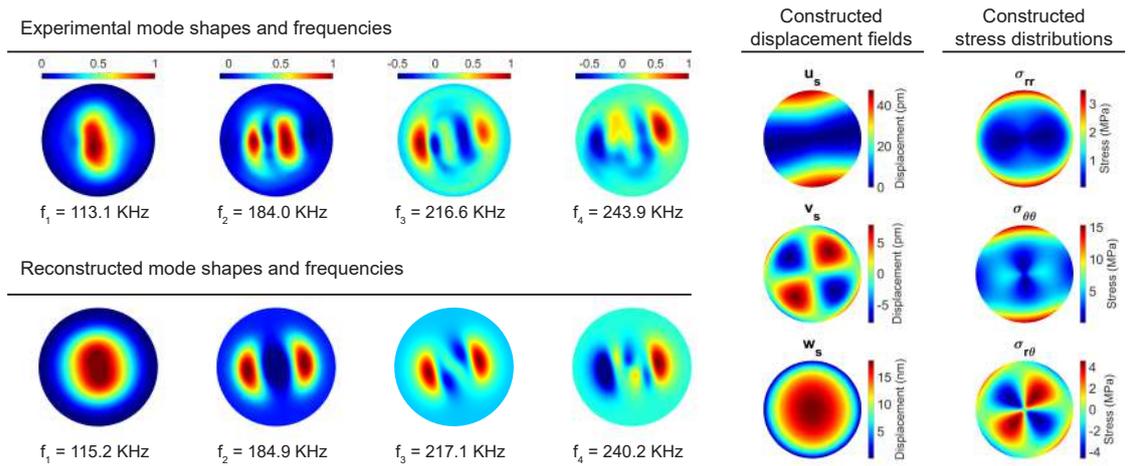

Figure S11: Estimation of stress distributions from experimental mode shapes for device D13.

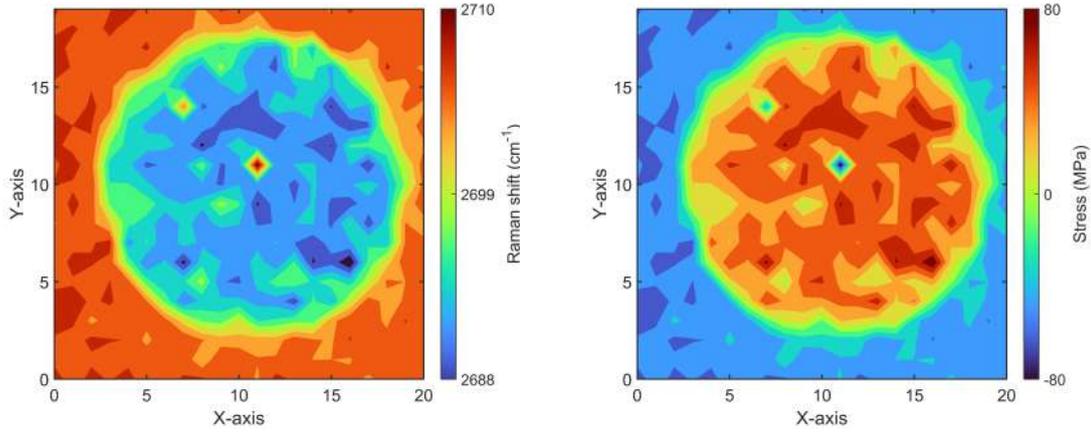

Figure S12: Measured Raman spectra of device D1 and the stress distribution extracted from the reported strain-dependent 2D peak.

Raman peak shift. The predicted stress range (difference between the maximum and minimum stress) obtained through Raman measurements is 16 times larger than the range predicted by our method. Additionally, the limited spatial resolution of Raman measurements is evident, with no clear stress distribution observable in the membrane, unlike the distinct distribution obtained through our method, as illustrated in figure 3 of the main text. Consequently, our proposed method offers a significantly more accurate approach for quantifying the stress distribution in suspended graphene membranes.



  \clearpage
  \restoregeometry
}

\end{document}